\documentclass[%
superscriptaddress,
amsmath,amssymb,
aps,
prb,
sort&compress,numbers
]{revtex4-2}
\usepackage{float}
\usepackage{amsmath}
\usepackage{amssymb}
\usepackage{tikz} 
\usepackage{wasysym}
\usepackage{graphicx}
\usepackage{xcolor}
\usepackage{bm}
\usepackage{pdfpages}

\makeatletter
\AtBeginDocument{\let\LS@rot\@undefined}
\makeatother

\usepackage{hyperref}
\usepackage{xr}

\begin{document}

\title[Benchmarking study of MLIPs]{Benchmarking of machine learning interatomic potentials for reactive hydrogen dynamics at metal surfaces}

\author{Wojciech G. Stark}%
\address{Department of Chemistry, University of Warwick, Gibbet Hill Road, Coventry CV4 7AL, United Kingdom}

\author{Cas van der Oord}%
\address{Department of Engineering, University of Cambridge, United Kingdom}

\author{Ilyes Batatia}%
\address{Department of Engineering, University of Cambridge, United Kingdom}

\author{Yaolong Zhang}%
\address{Department of Chemistry and Chemical Biology, Center for Computational Chemistry, University of New Mexico, Albuquerque, New Mexico 87131, USA}

\author{Bin Jiang}%
\address{Key Laboratory of Precision and Intelligent Chemistry, Department of Chemical Physics, University of Science and Technology of China, Hefei, Anhui, China}
\address{Hefei National Laboratory, University of Science and Technology of China, Hefei 230088, China}

\author{Gábor Csányi}%
\address{Department of Engineering, University of Cambridge, United Kingdom}
    
\author{Reinhard J. Maurer}%
\email{r.maurer@warwick.ac.uk}
\address{Department of Chemistry, University of Warwick, Gibbet Hill Road, Coventry CV4 7AL, United Kingdom}
\address{Department of Physics, University of Warwick, Gibbet Hill Road, Coventry CV4 7AL, United Kingdom}


\begin{abstract}
    Simulations of chemical reaction probabilities in gas surface dynamics require the calculation of ensemble averages over many tens of thousands of reaction events to predict dynamical observables that can be compared to experiments. At the same time, the energy landscapes need to be accurately mapped, as small errors in barriers can lead to large deviations in reaction probabilities. This brings a particularly interesting challenge for machine learning interatomic potentials, which are becoming well-established tools to accelerate molecular dynamics simulations. We compare state-of-the-art machine learning interatomic potentials with a particular focus on their inference performance on CPUs and suitability for high throughput simulation of reactive chemistry at surfaces. The considered models include polarizable atom interaction neural networks (PaiNN), recursively embedded atom neural networks (REANN), the MACE equivariant graph neural network, and atomic cluster expansion potentials (ACE). The models are applied to a dataset on reactive molecular hydrogen scattering on low-index surface facets of copper. All models are assessed for their accuracy, time-to-solution, and ability to simulate reactive sticking probabilities as a function of the rovibrational initial state and kinetic incidence energy of the molecule. REANN and MACE models provide the best balance between accuracy and time-to-solution and can be considered the current state-of-the-art in gas-surface dynamics. PaiNN models require many features for the best accuracy, which causes significant losses in computational efficiency.  ACE models provide the fastest time-to-solution, however, models trained on the existing dataset were not able to achieve sufficiently accurate predictions in all cases.
\end{abstract}

\maketitle

\section{Introduction} \label{sec:intro}

    Hydrogen evolution and hydrogenation reactions play vital roles in heterogeneous catalysis~\cite{ertl_reactions_2008,sabbe_first-principles_2012,smith_current_2020,waugh_methanol_1992,grabow_mechanism_2011,behrens_active_2012}. By gaining a deeper understanding of such fundamental processes at the atomic scale, we pave the way for designing improved catalysts. This is why simulations of explicit surface dynamics are crucial. Achieving the desired level of fidelity can be achievable through the use of (\textit{quasi}-)classical trajectory simulations, which take into account adsorbate and substrate motion. Furthermore, to predict experimentally measurable reaction probabilities, ensemble averages over many tens of thousands of events are required, which is unfeasible with on-the-fly \textit{ab-initio} molecular dynamics (MD) simulations based on electronic structure methods, such as density functional theory (DFT). Interatomic potentials are commonly employed as efficient alternatives to \textit{ab-initio} MD.

    The first generation of such interatomic potentials for gas-surface dynamics was based on the corrugation reducing procedure (CRP)~\cite{busnengo_representation_2000,busnengo_surface_2005}, modified Shepard interpolation~\cite{ischtwan_molecular_1994,thompson_molecular_1997} or the permutation invariant polynomials (PIPs)~\cite{braams_permutationally_2009,bowman_high-dimensional_2011}, which turned out to be successful for many systems~\cite{salin_theoretical_2006,jiang_six-dimensional_2014,lozano_adsorption_2009,tchakoua_toward_2019,diaz_dynamics_2010,crespos_multi-dimensional_2003,crespos_application_2004}. However, these methods could not include many degrees of freedom, so surface atom motion was typically not explicitly included and instead accounted for approximately~\cite{mondal_thermal_2013}. More recently proposed CRP models, known as dynamic corrugation models (DCM) allowed the inclusion of surface degrees of freedom by combining CRP with embedded atom method (EAM) potentials~\cite{smits_beyond_2021}, the accuracy of which is not yet fully assessed.
    
    More recently developed machine learning interatomic potentials (MLIPs) based on atom-centered representations enable the straightforward inclusion of all degrees of freedom of the system and are now commonly employed for a variety of material classes.~\cite{deringer_machine_2019, unke_machine_2021, kulik_roadmap_2022}
    Recent improvements in MLIPs brought about excellent computational efficiency without loss of accuracy of the underlying first-principles data. However, tackling chemical dynamics at surfaces often continues to pose challenges. For instance, predicting the sticking probability of molecular hydrogen on copper as a function of kinetic incidence energy and the vibrational initial state with robust statistical sampling requires many millions of MD trajectories in total.~\cite{zhu_unified_2020} An MLIP with an inference time of 10~ms per atom per MD time step equates to only 1~s per time step for a 100-atom system, but to 11,574~days for 1~million trajectories in sequence (assuming 1,000 time steps each). This is still unfeasible despite the orders of magnitude speed-up compared to DFT. Therefore gas surface dynamics simulations provide a difficult but interesting challenge for MLIPs.
    
    To simulate reactions using ML methods, descriptors are used to represent the atomic environment. To satisfy symmetry requirements, descriptors have to capture the rotational, permutational, and translational invariance of the energy. Most popular high-dimensional descriptors that meet these requirements include atom-centered symmetry functions (ACSF)~\cite{behler_atom-centered_2011} and the smooth overlap of atomic positions (SOAP)~\cite{bartok_representing_2013}. Although both proved to be very successful in modeling atomic interactions, they are both limited to three-body descriptions of the local environment and therefore are incomplete:  they are not be able to distinguish between certain atomic environments~\cite{pozdnyakov_incompleteness_2020}. Recently, Drautz introduced the atomic cluster expansion (ACE)~\cite{drautz_atomic_2019,drautz_atomic_2020}, which provides an efficient many-body description of the atomic environment to arbitrary body order. 
    Simultaneously, equivariant features in the context of message passing networks were demonstrated to improve the expressiveness and completeness of representations by including vectorial or tensorial representations.~\cite{schutt_equivariant_2021,batatia_design_2022, batzner_e3-equivariant_2022,musaelian_learning_2023}

    Message-passing neural networks (MPNNs) are gaining increased  popularity~\cite{gilmer_neural_2017,schutt_quantum-chemical_2017,schutt_equivariant_2021,batatia_mace_2022} due to their ability to learn the representations of atomic geometry by iterative exchange of messages between nodes (atoms). An example of such an approach is the polarizable atom interaction NN (PaiNN)~\cite{schutt_equivariant_2021}, which combines this approach with the inclusion of equivariant features, thereby allowing even more information to be embedded in the descriptors. Another very successful example of MPNN is the recursively embedded atom neural network (REANN)~\cite{zhang_physically_2021}, in which embedded atom densities are used as atomic descriptors, and the relevant coefficients are learned through the message-passing process. REANN passes three-body messages, and thus can reach a complete description of the atomic environment after 3 rounds of message passing~\cite{rose2023iterations_WL}. Another popular MPNN is MACE~\cite{batatia_mace_2022}, which similarly combines message passing with equivariant features, but instead of passing just two-body messages (e.g. PaiNN), MACE allows passing arbitrary higher body order messages, through the utilization of the ACE framework. In a typical setting, 4-body features are constructed, and a complete representation is reached after just 1 iteration of message passing~\cite{rose2023iterations_WL}. Note that in the present case studied in this paper, we find that three-body features are sufficient for MACE also.


    Recently, Stark~\textit{et~al.}~\cite{stark_machine_2023} have performed an adaptive sampling of training data based on the widely-used specific reaction parameter (SRP)-based functional~\cite{gonzalez-lafont_direct_1991,diaz_chemically_2009} to construct  MLIPs for H\textsubscript{2} chemistry on various facets of copper. Specifically, they have compared the difference in the performance of an equivariant (PaiNN) and a non-equivariant MPNN (SchNet) for the simulation of reactive scattering probabilities, with PaiNN MLIPs outperforming SchNet-based MLIPs in accuracy and data efficiency. Jiang and coworkers have previously presented a similar model based on the EANN.~\cite{zhu_unified_2020} While the equivariant MPNN model and the EANN model were both able to accurately predict dynamical sticking probabilities, force evaluations with the MPNN model, PaiNN, are computationally expensive, which makes statistical convergence of dynamical reaction probabilities computationally challenging.
    
    In this study, we systematically assess the ability of commonly used MLIP models to simultaneously offer high accuracy, reliability, and efficiency of force evaluation for gas-surface dynamics simulations.  We compare different architectures, such as PaiNN, ACE (linear), MACE, and REANN, and provide recommendations for their uses in chemical reaction dynamics at surfaces by analyzing the accuracy of the models, smoothness of the potential energy surface (PES), and evaluation times. In all cases, a trade-off needs to be achieved between accuracy and inference efficiency during hyperparameter optimization.
    

\section{Methods} \label{sec:methods}

    \subsection{Descriptors and interatomic potentials} \label{sec:methods_ips}
        \paragraph{Atomic Cluster Expansion} \label{sec:methods_ips_ace}
            ACE provides a framework to construct highly efficient interatomic potentials that achieve high body order representation. Contrary to PIPs, ACE does not require a direct summation over all atom tuples.
            The atomic basis $A_{z_{i},znlm}$ is defined as a projection of the neighborhood density $\rho_i^{z}$ of all atoms of chemical element $z$ around atom $i$ (where atom $i$ belongs to the chemical element $z_i$) onto the one-particle basis functions $\phi_{nlm}^{z_{i}z_{j}}$
            \begin{equation}
                A_{z_{i},znlm} = \left\langle\rho_{i}^{z}|\phi_{nlm}^{z_{i}z}\right\rangle = \sum_{\substack{j\\
                  \forall z_j=z}} \phi_{nlm}^{z_{i}z_{j}}(r_{ji})  ,
            \end{equation}
            where the one-particle basis functions are summed over atoms $j$, for which $z_{j}=z$. The $z_{i},z_{j}$ are elements $z$ of central $i$ and neighboring $j$ atoms, and the density can be described as
            \begin{equation}
                \rho_{i}^{z}(\pmb{r})=\sum_{\substack{j \neq i \\ \forall z_j =z}}\delta(\pmb{r}-\pmb{r}_{ji}) ,
            \end{equation}
            and the one-particle basis function can be defined as a product of a radial function $R_{nl}^{z_{i}z_{j}}$ and spherical harmonics $Y_{l}^{m}$,
            \begin{equation}
                \phi_{nlm}^{z_{i}z_{j}}(\pmb{r})=R_{nl}^{z_{i}z_{j}}(r)Y_{l}^{m}(\pmb{\hat{r}}) .
            \end{equation}
            To obtain permutation-invariant basis functions, products of atomic basis functions are constructed 
            \begin{equation}
                A_{z_{i}\pmb{\upsilon}}=\prod^{\nu}_{t=1} A_{z_{i}\upsilon_{t}},
            \end{equation}
            where $\upsilon = znlm$, $\pmb{\upsilon}=(\upsilon_{1},...,\upsilon_{\nu})$ and $\nu$ refers to correlation order, corresponding to a ($\nu$+1)-body basis function. 
            Isometry invariant basis $B_{z_{i}\eta_{\nu}}$ functions are created from the non-rotationally-invariant  basis functions $A_{z_{i}\pmb\upsilon}$ 
            \begin{equation}
                B_{z_{i}\eta_{\nu}}=\sum_{\pmb{\upsilon}} C_{\eta_{\nu}\pmb{\upsilon}}A_{z_{i}\pmb{\upsilon}} ,
            \end{equation}
            where $C_{\eta_{\nu}\pmb{\upsilon}}$ are Clebsch-Gordan coefficients, and all the allowed invariant contractions of $A_{z_{i}\pmb{\upsilon}}$ are enumerated by $\eta_{\nu}$. The site energy can be then defined as
            \begin{equation}
                E_{i}=\sum c_{z_{i}\pmb{\upsilon}}B_{z_{i}\eta_{\nu}} ,
            \end{equation}
            where $c_{z_{i}\pmb{\upsilon}}$ coefficients are free model parameters optimized in a linear fitting procedure.

        \paragraph{Message-passing neural networks} \label{sec:methods_ips_mpnns}
            PaiNN and MACE are MLIPs based on MPNNs. In MPNNs, each atom (node) is connected with edges to all neighboring atoms within a specified cutoff distance, to create graphs embedded in 3-dimensional Euclidean space. The MPNN graph layout enables the passing of information between atoms through a series of message passing-update steps, creating a representation that indirectly carries details about atoms from outside of the cutoff distance and can encode many-body interactions. This process can be described as
            \begin{equation} \label{eqn:message}
                \pmb{m}^{t+1}_{i} = \sum_{j\in \mathcal{N}(i)} \textbf{M}_{t}(\pmb{s}^{t}_{i},\pmb{s}^{t}_{j},\pmb{e}_{ij}),
            \end{equation}
            \begin{equation} \label{eqn:h_eq}
                \pmb{s}^{t+1}_{i} = \textbf{U}_{t}(\pmb{s}^{t}_{i},\pmb{m}^{t+1}_{i}),
            \end{equation}
            where $\pmb{m}^{t+1}_{i}$ is a message created by summing over the scalar atomic features $\pmb{s}_i$ of nodes (atoms) $i,j$ at step $t$, connected by edge features $\pmb{e}_{ij}$ (usually interatomic distances $r_{ij}$). $\textbf{M}_{t}$ and $\textbf{U}_{t}$ can be linear or nonlinear message and update functions, respectively~\cite{gilmer_neural_2017}.
            
            Recently, equivariant (vectorial) features embedded in MPNNs proved to provide significant improvement in the data efficiency and accuracy of the models~\cite{schutt_equivariant_2021, batzner_e3-equivariant_2022,musaelian_learning_2023,batatia_design_2022}. To include such features in the message passing scheme, the message in Eq.~\ref{eqn:message} is adapted to
            \begin{equation} \label{eqn:message_eq}
                \vec{\pmb{m}}^{t+1}_{i} = \sum_{j\in \mathcal{N}(i)} \vec{\textbf{M}}_{t}(\pmb{s}^{t}_{i},\pmb{s}^{t}_{j},\hat{\pmb{v}}^{t}_{i},\hat{\pmb{v}}^{t}_{j},\pmb{e}_{ij}),
            \end{equation}
            where vectorial representations $\hat{\pmb{v}}$ are additionally included.

        \paragraph{Embedded Atom Neural Networks} \label{sec:methods_ips_reann}
            Similarly to EAM, EANN makes use of the electron density of an atom embedded in the material to predict the atomic contributions to the energy.~\cite{zhang_embedded_2019} This is done by evaluating Gaussian-type orbitals centered at each atom,
            \begin{equation}
                \varphi ^{n}_{l_{x}l_{y}l_{z}}(\pmb{\hat{r}_{ij}})=(x_{ij})^{l_{x}}(y_{ij})^{l_{y}}(z_{ij})^{l_{z}} \cdot \exp\left[-\alpha(r_{ij}-r_{s})^{2}\right]f_{c}(r_{ij}),
            \end{equation}
            where $r_{ij}=|\pmb{\hat{r}_{ij}}|$ is an interatomic distance between central atom $i$ and neighbor atom $j$. Similarly, $x_{ij}=x_{j}-x_{i}$, $y_{ij}=y_{j}-y_{i}$ and $z_{ij}=z_{j}-z_{i}$.
            $l=l_{x}+l_{y}+l_{z}$ is orbital angular momentum, $\alpha$ and $r_{s}$ are hyperparameters that determine the center and the width of the radial Gaussian function and $f_{c}(r_{ij})$ is a cutoff function.
            The embedded atom density is then calculated by taking a linear combination of the Gaussian-type orbitals of neighboring atoms 
            \begin{equation}
                \rho_{i}^{n}=\sum^{l_{x}+l_{y}+l_{z}=l}_{l_{x},l_{y},l_{z}}\frac{l!}{l_{x}!l_{y}!l_{z}!}\left[\sum^{N_{\mathrm{atom}}}_{j=1}c_{j}\varphi^{n}_{l_{x}l_{y}l_{z}}(\pmb{\hat{r}}_{ij})\right]^{2},
            \end{equation}
            where $c_{j}$ is an expansion coefficient of an orbital of atom $j$ and $N_{\mathrm{atom}}$ is the number of neighboring atoms within the cutoff radius distance. 
            
            In REANN, different atomic orbitals are mixed with angular momentum before creating the molecular orbital, in such a way that
            \begin{equation}
                \rho_{i}^{n}=\sum^{L}_{l=0}\sum^{l_{x}+l_{y}+l_{z}=l}_{l_{x},l_{y},l_{z}}\frac{l!}{l_{x}!l_{y}!l_{z}!}\left[\sum^{N_{\mathrm{wave}}}_{m=1}d^{n}_{m} \sum^{N_{\mathrm{atom}}}_{j\neq i}c_{j} \varphi^{m}_{l_{x}l_{y}l_{z}}(\pmb{\hat{r}}_{ij})\right]^{2},
            \end{equation}
            where $L$ is the maximum orbital angular momentum of primitive Gaussian-type orbitals (GTOs), $N_{\mathrm{wave}}$ is the number of primitive GTOs and $d_m^n$ is a contraction coefficient of the m-th primitive GTO for generating the n-th contracted GTO. The embedded atom density is a three-body descriptor when L\textgreater0, which incorporates radial and angular information implicitly. 
            
        \subsection{Computational details} \label{sec:methods_ips_settings}
            \subsubsection{Database and ab initio calculations}
                The database includes 4230 datapoints, with 1685 Cu-only surface structures sampled at different temperatures (300~K, 600~K, and 900~K), and 2545 H\textsubscript{2}/Cu structures. The database contains structures of four different Cu facets, namely Cu(111), (100), (110), and (211). We employed 3$\times$3, 6-layered slabs for all the surfaces in the database. To create a reliable database, we performed adaptive sampling based on iteratively refining a SchNet MLIP. This involved performing dynamics simulations, picking structures that are not well represented by our models trained on the current database, adding them, and retraining the models.
                More details about the database generation can be found in Ref.~\cite{stark_machine_2023}. As the training database was generated with SchNet, none of the models used in this study (PaiNN, MACE, ACE, REANN) have been involved in the sampling of the dataset.
                
                For DFT calculations we employed a specific reaction parameter (SRP) functional~\cite{nattino_effect_2012} containing 52\% of PBE~\cite{perdew_generalized_1996} and 48\% of RPBE functional~\cite{hammer_improved_1999} (SRP48), which is known to correctly predict the dissociation barrier.~\cite{diaz_chemically_2009,sementa_reactive_2013,marashdeh_surface_2013,kroes_vibrational_2017,cao_hydrogen_2018,smits_quantum_2023,stark_machine_2023}. We employed a k grid of 12$\times$12$\times$1 and a ``tight'' default basis set within the FHI-aims~\cite{blum_ab_2009} all-electron electronic structure code.
                
                Minimum energy paths with MLIPs were obtained using climbing image nudged elastic bands~\cite{henkelman_climbing_2000} with 50 images and maximum force along the path of 0.01~eV/$\textrm{\AA}$.
            \subsubsection{Model parameters} \label{sec:methods_ips_settings_painn} 

                The final optimized PaiNN model includes 5 interaction layers and 128 features in message passing. Additionally, a cutoff distance of 4~$\textrm{\AA}$ and loss function energy and force weights of 0.05 and 0.95 respectively, were used in the training process. The models were trained using SchNetPack code~\cite{schutt_equivariant_2021,schutt_schnetpack_2023} (\url{https://github.com/atomistic-machine-learning/schnetpack}), version 2.0.0.

                For the ACE models, a cutoff distance of 6~$\textrm{\AA}$ was employed. Additionally, a correlation order of 4 with the corresponding polynomial degrees of 18, 12, 10, and 8 were set. Energy and force weights of 0.5 and 0.5, respectively, were used in the combined loss function. The models were trained using the ACEpotentials.jl code~\cite{witt_acepotentialsjl_2023, drautz_atomic_2019, dusson_atomic_2022} (\url{https://github.com/ACEsuit/ACEpotentials.jl}), version 0.6.3. In the following sections, we will address the ACEpotentials.jl models as ACE models.

                Cutoff distance of 5~$\textrm{\AA}$, correlation order of 2 and 2 interactions were used in the final MACE models. Initial loss function energy and force weights of 5 and 95, respectively, were used in the training process. After 1200 epochs the weights were changed to 1000 and 1 for energies and forces, respectively. Model size (\verb|hidden_irreps|) in the final model was set to \textit{16x0e}, so equivariant message passing was found to be not necessary to obtain high accuracy results in the present case. The models were trained using the MACE code ~\cite{batatia_mace_2022,batatia_design_2022} (\url{https://github.com/ACEsuit/mace}), version 0.2.0. MACE-JAX models were trained using the MACE-JAX code (\url{https://github.com/ilyes319/mace-jax.git}).
                
                For the final REANN models, a cutoff distance of 5~$\textrm{\AA}$ was employed. The NN structure for atomic energies was set to \verb|16|$\times$\verb|16|$\times$\verb|16| and for orbital coefficients to \verb|16|$\times$\verb|16|$\times$\verb|16|. The number of message-passing loops was set to 3. The number of primitive GTOs was set to 9. The loss function energy weight stayed at 0.1 for the entire training process. The initial weight for force was 50 and was reduced throughout the training process to the final value of 0.5. The models were trained using REANN code~\cite{zhang_reann_2022,zhang_physically_2021} (\url{https://github.com/zhangylch/REANN}), version 0.1. In Section~\ref{sec:results_time_efficiency} we additionally show the performance of the newest REANN (version 1.0), with Fortran-based neighbor list code, which requires an additional compilation step for the Fortran code.

                For model error evaluation, we use two metrics, mean absolute error (MAE) and root-mean-square error (RMSE). The latter is more sensitive to outliers. The final settings of all the models were established by a cross-validation optimization of model parameters. 
                
                During this study, we identified non-zero spin structures in the previously published dataset, in which one of the hydrogen atoms is adsorbed and the other atom is detached from the surface (examples are shown in Supplementary Figure~S1). Only 10 such structures were identified in our dataset (4230 data points), however, the structures caused inflated energy RMSEs, particularly for MACE models due to the resulting inconsistency for the energy of isolated atoms. The presence of these structures did not impact the learning abilities of the models (Supplementary Figure~S2), which we discuss further in Section~\ref{sec:results_time_efficiency}. These are unusual geometries that will not be encountered in molecular scattering dynamics under relevant conditions. Therefore, the structures were removed from the test sets of all models, enabling a more reliable comparison.

                More details about all the models, including the parameter optimization, are available in the Supplementary Information.
                
        \subsubsection{Molecular dynamics simulations} \label{sec:methods_md_settings}
            All molecular dynamics simulations, including preparation of initial conditions and analysis of results, were performed using NQCDynamics package~\cite{gardner_nqcdynamicsjl_2022} (\url{https://github.com/NQCD/NQCDynamics.jl}, version 0.13.4).
            
            The output of MD simulations of hydrogen molecule adsorption on metal surfaces can be used for calculating sticking (reaction) probabilities. The simulations are initiated with the hydrogen molecule at 7~$\textrm{\AA}$ above the surface and vibrational and rotational initial conditions for H\textsubscript{2} are established using Einstein-Brillouin-Keller (EBK) method~\cite{larkoski_numerical_2006} for several normal incidence energies and rovibrational states with randomly chosen polar and azimuthal angles. Simulations at 0~K surface temperatures were initiated using a DFT-relaxed slab with initial velocities based on the Maxwell-Boltzmann distribution. Simulations at finite temperatures were initialized with surface positions received from a random sampling of surface-only Langevin MD output structures at set temperatures. Lattice expansion is included based on the relation between lattice constant and surface temperature. This is done by running NPT simulations at 9 different temperatures, ranging from 200 to 1000~K (Supplementary Figure~S3), and fitting a linear function, to determine the lattice constant in 925~K. This is done with ACE, MACE, and PaiNN models, however, the outcome is roughly the same for all the codes. We did not include results obtained with REANN, due to the fact that NPT simulations with REANN were unstable. The use of MLIPs, instead of direct DFT calculations, to evaluate the lattice constants can be justified by their reliability in predicting the energies at different lattice constants, as shown in Supplementary Figure~S4. The final lattice constant used for the simulations in 925~K is 3.71~$\textrm{\AA}$ (3.65~$\textrm{\AA}$ at 0~K). Sticking probabilities were calculated using data averaged over 10,000 trajectories for every model, surface facet, rovibrational state, and incidence energy reported in this study. The maximum simulation time of every trajectory was 3~ps with a time step of 0.1~fs unless special conditions were met, such as the distance between hydrogen atoms when adsorbed at the metal surface being above 2.25~$\textrm{\AA}$ (counted as dissociation event) or the distance between hydrogen molecule and top surface atoms being above 7.1~$\textrm{\AA}$ (counted as scattering event).

\section{Results and discussion}\label{sec:results}

    \subsection{Model parameter optimization and learning behavior} \label{sec:results_opt}
        We performed cross-validation optimization for all of the models used in this study. The optimization of the most impactful parameters, including energy, all-atom force, and H-atom-only force test RMSEs, as well as the energy evaluation times, is shown in Supplementary Figures~S5-S9. 
        
        For the PaiNN models, we carried out an optimization over the number of interaction layers, features, and the cutoff distance (Supplementary Figure~S5). The convergence with respect to the interaction layers and features is relatively slow (reaching convergence at 7 interaction layers and 512 features), however, satisfactory levels of energy and force RMSEs can be achieved already with 5 interaction layers and 128 features, giving a much-needed tenfold improvement in efficiency over the energy evaluation speed obtained with 7 interaction layers and 512 features. Cutoff distance converges between 3 and 4~$\textrm{\AA}$, thus, we use 4~$\textrm{\AA}$ in our final model. The low RMSEs obtained for such a low cutoff distance are possible due to the message passing included in PaiNN (as well as REANN and MACE) that allows iterative passing of information from outside of the cutoff to the feature vectors of the atoms inside of it~\cite{stark_machine_2023}. The final PaiNN models include 5 interaction layers, 128 features, and a cutoff of 4~$\textrm{\AA}$.
        
        The optimization of REANN model parameters is depicted in Supplementary Figure~S6~and~S7, including the convergence of the NN structure (number of neurons and hidden layers), cutoff distance, number of primitive GTOs (nwaves), and maximum angular momentum (L). The convergence of the number of neurons using two hidden layers is slow and concludes at 256 neurons per layer. Convergence is achieved much faster for 3 hidden layers (already at 16$\times$16$\times$16) and allows obtaining much lower energy evaluation times at the same accuracy as 256$\times$256 (left side of Supplementary Figure~S6). Cutoff distance values converge quickly, already at 4~$\textrm{\AA}$ (right side of Supplementary Figure~S6), however, when simulating sticking probabilities, we noticed an improvement with a cutoff of 5~$\textrm{\AA}$. Optimization of the number of primitive GTOs used for the training of REANN models is included on the left side of Supplementary Figure~S7. The convergence is relatively slow. We see more stable behavior when using 7 or more primitive GTOs. The impact of using more primitive GTOs on evaluation time is low. We therefore chose 9 primitive GTOs in our final models. Lastly, the optimization of maximum angular momentum (L), shown on the right side of Supplementary Figure~S7, provides proof that the maximum angular momentum can be kept low without any loss of accuracy. For the final REANN models, we used 16 neurons with 3 hidden layers for the NN, a cutoff of 5~$\textrm{\AA}$, 7 primitive GTOs, and a maximum angular momentum of 1.

        The ACE parameters, such as correlation order, corresponding polynomial degrees, and cutoff distance, were also optimized (Supplementary Figure~S8). The convergence of correlation order, together with the corresponding polynomial degrees is shown on the left side of Supplementary Figure~S8. To achieve the lowest RMSEs, a correlation order of at least 3 is required, as well as a polynomial degree of at least 18. We noticed that it is beneficial to use varying (decreasing) polynomial degrees for each correlation order, as it allows the best combination of accuracy and efficiency. The convergence of cutoff distance is slower than for the NN-based codes, as expected, and a cutoff of at least 5.5~$\textrm{\AA}$ is required for the best accuracy. For the final models, we use ACE models with a correlation order of 4 with the polynomial degrees of 18, 12, 10, and 8, as well as the cutoff distance of 6~$\textrm{\AA}$.
        
        The model size and the cutoff distance employed in our MACE models were optimized and plotted in Supplementary Figure~S9. The convergence with respect to the model size (left side of Supplementary Figure~S9) is relatively fast and is obtained already for \textit{16x0e}. The equivariant features provide significant improvement for very small model sizes, such as \textit{2x0e}, however, there are no significant differences between \textit{16x0e} and \textit{16x0e 16x1o}. Therefore in this case using models with strictly invariant message passing can be beneficial in terms of computational efficiency. The convergence of the cutoff, shown on the right side of Supplementary Figure~S9 is, similar to PaiNN and REANN, reached at 4~$\textrm{\AA}$. However, just like with REANN models, we see further improvements when calculating reaction probabilities with a cutoff distance of 5~$\textrm{\AA}$. For the final MACE models employed in this study, invariant \textit{16x0e} features and a cutoff of 5~$\textrm{\AA}$ were used.

        \begin{figure}
            \centering
            \includegraphics[width=1.0\linewidth]{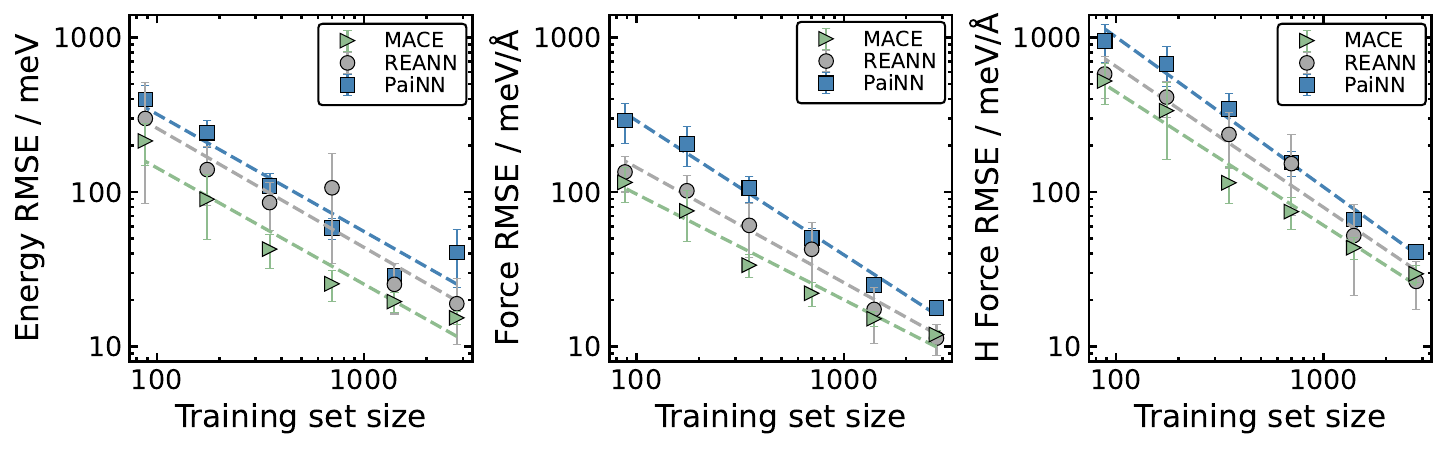}
                \caption{\textbf{Learning curves for MACE, REANN, and PaiNN models.} Log-log plot of the averaged test set predictive error (RMSE) in energy, force, and hydrogen-atom-only force as a function of training set size for all NN-based models employed in this study. The shown RMSE values are averaged over 5 cross-validation splits, differing by training and validation sets. Error bars correspond to standard deviations between the RMSEs obtained over all splits.}
            \label{fig:learning_curves}
        \end{figure}

        Learning behavior for different training set sizes, with respect to energy, force, and H-atom-only force, of all of the NN-based codes (MACE, REANN, and PaiNN) is shown in Fig.~\ref{fig:learning_curves}. MACE models are the most efficient in learning both energies and forces, especially for smaller training sizes, however, the gradient in learning curves seems to change at larger training dataset sizes beyond roughly 700 data points and improve at a slower pace with more training data. This behavior is not as apparent in the learning curves that correspond to REANN and PaiNN, for which the behavior is mostly linear at least up until the data size shown. The general gradients of the energy and force learning curves are roughly the same for all the models. Of all the considered NN-based models, PaiNN is the least efficient in learning both energies and forces. Predictions made with REANN models are associated with the highest relative deviations among all the methods. All the models provide a satisfactory level of RMSEs for large training set sizes with energy RMSEs reaching levels below the chemical accuracy (1~kcal/mol $\approx$ 43~meV) and hydrogen atom force RMSEs reaching below 41~meV/$\textrm{\AA}$ for PaiNN models and even lower (30~meV/$\textrm{\AA}$) for REANN and MACE models. The improvement in learning all forces, as well as, hydrogen-only forces with more training data is similar for all the codes. The errors associated with hydrogen forces are significantly larger than the overall force errors. This may be due to the small number of hydrogen atoms contained in the slab models (56 atoms overall per unit cell with only 2 hydrogen atoms) and the more complex nature of learning the reaction barriers of hydrogen dissociation. Supplementary Figure~S10 reports the learning curves based on energy and force MAE. The energy learning curves of REANN and PaiNN based on MAEs are close together and force learning curves obtained with REANN reach the same accuracy level as MACE models. 

    \subsection{Computational efficiency of the models}   \label{sec:results_time_efficiency}

\begin{figure}
            \centering
            \includegraphics[width=1.0\linewidth]{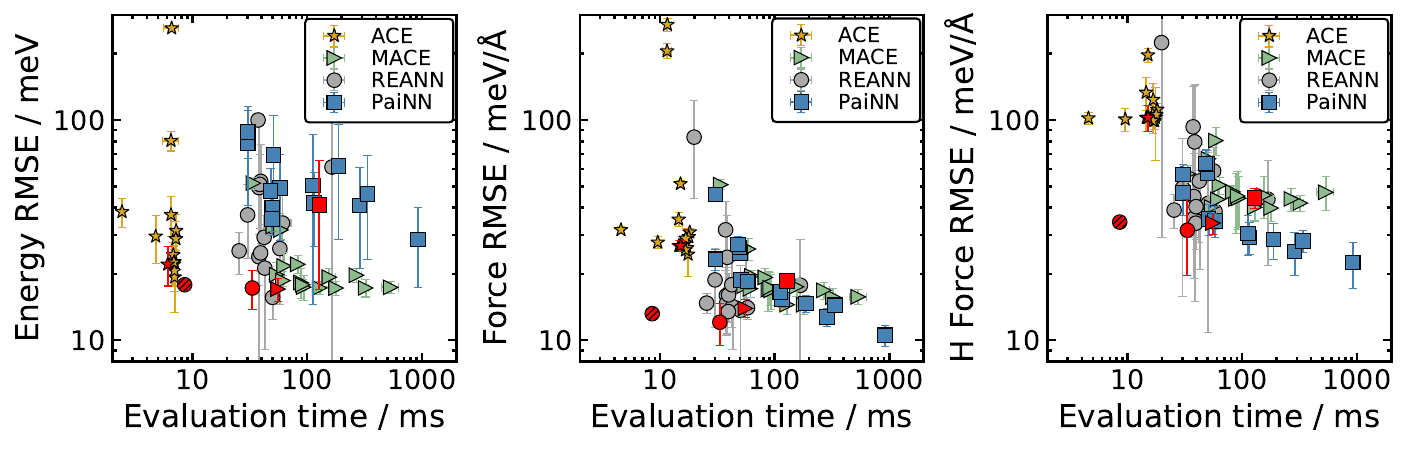}
                \caption{\textbf{Energy and force (all atoms and H only) test RMSEs in meV and meV/$\textrm{\AA}$ plotted against evaluation time in milliseconds on CPUs.} We include the results obtained with models trained for cross-validation, using PaiNN (blue squares), REANN (gray circles), MACE (green triangles), and ACE (yellow stars) models. Data points and error bars correspond to the RMSE average and standard deviation over 5-fold cross-validation splits. Red-colored data points correspond to the final models discussed in Section \ref{sec:results_errors}. The striped circle corresponds to the new REANN version, which uses a Fortran-based neighbor list calculator. The energy and force evaluation times were calculated using a single AMD EPYC 7742 (Rome) 2.25~GHz CPU processor core.}
            \label{fig:error_vs_time}
        \end{figure}

        In the previous section, we discussed the optimization of model parameters performed using a database obtained from Ref.~\cite{stark_machine_2023}, with all of the methods used in this study, namely, ACE, MACE, REANN, and PaiNN. Figure~\ref{fig:error_vs_time} plots the energy and force test RMSEs (including both all-atoms and H-atom-only force test RMSEs) of 15 different trained models for each method against their corresponding CPU evaluation times. 
        We chose to evaluate all models on CPUs only, because this allows a direct comparison, and also because a large number of CPU cores were available and can be efficiently utilized for the high throughput evaluation of many short trajectories,  which is the specific feature of the computational task discussed here. Note that the MACE code is specifically written for GPUs (utilizing either PyTorch or JAX) and its efficiency on CPUs is highly suboptimal. 
        The models shown in the figure were created during cross-validation and hyperparameter optimization (discussed in the previous section), thus they explore a wide range of settings. Additionally, the optimized models trained using the final settings are also plotted in red, with RMSEs evaluated over the same training-test splits as for the other models. This visualization allows us to have a better perspective on the trade-off between prediction accuracy and evaluation performance.
        The striped circles represent the REANN models which are evaluated using the new Fortran-based neighbor list code (REANN-Fort), as opposed to the Python-based neighbor list implementation (REANN-Py) depicted with solid markers.

        The figure contains test errors obtained with different models, in which we exclude previously identified outliers with non-zero spin  (example structures shown in Supplementary Figure~S1). We identified 10 such structures in the entire dataset of 4230 structures. Typically, only 0-2 of such structures would end up in the split-off test set. In Supplementary Figure~S2, we included an analogous plot to Fig.~\ref{fig:error_vs_time} only for the optimized models where the RMSEs are compared for the original test set (orange markers), the original test set without high-spin structures (red markers) and finally, we show the data points corresponding to a new model trained without such structures included in either training or test sets (black markers). The final test errors obtained using ACE, REANN, and PaiNN models are almost identical with and without non-zero spin structures in the test set and very close (within the uncertainty) to the errors obtained with the models trained without outliers. Force prediction errors obtained with MACE models show the same behavior, however, there is a large discrepancy between the energy RMSEs obtained for the test set with and without the mentioned structures. The discrepancy is caused by only 2 such structures in the test set. Furthermore, there is no significant change in the RMSE prediction for the primary model without the non-zero-spin structures in the test set and the newly trained model that excludes such structures in the entire dataset. Such discrepancy in energy RMSEs, without a corresponding discrepancy in force errors leads us to conclude that this is most likely caused by the difference in the treatment of atoms with respect to their reference energies (which for MACE is set to be the energy of the isolated atoms). The MACE model correctly infers that the atomic energy of the outliers should differ, which leads to the increased errors caused by the use of the same atomic energies for all H atoms. Such non-zero spin structures are not expected in dissociative chemisorption, which is why we excluded them from the test sets of all the models.
        
        The lowest energy RMSEs in the cross-validation were obtained with MACE, REANN, and ACE models, falling even below 20~meV using the most successful models. All these methods converge at a similar energy RMSE value, possibly reaching the limit of accuracy that can be reached for the chosen numeric convergence settings of the underlying DFT data. PaiNN models fail to approach this level of accuracy regardless of model hyperparameters and only reach RMSEs of roughly 30~meV with the most accurate models. Any further model improvements (e.g. increasing model size or cutoff distance) did not yield further prediction improvements. The spread of energy RMSEs over the five cross-validation data splits (represented by standard deviations) is the highest for the PaiNN and REANN models and in some cases reaches over 40~meV. This indicates that the models are more sensitive to the distribution of data points that they are presented with. ACE models achieve the lowest energy evaluation times on CPUs. For the most accurate ACE models, energy evaluations reach 6-7~ms per single evaluation. The second most efficient models were obtained with REANN-Fort, reaching the evaluation speeds of roughly 8-9~ms per single evaluation. REANN-Py, MACE and PaiNN reach similar energy evaluation times of roughly 30-50~ms per single evaluation. Taking into consideration both energy prediction accuracy and evaluation speed on CPUs, the models can be ranked as follows (starting with the most successful): ACE, REANN-Fort, REANN-Py, MACE, and PaiNN.

        PaiNN, MACE, and REANN models reach similarly low all-atom and H-atom-only force RMSEs at roughly 10-15~meV/$\textrm{\AA}$ and 25-35~meV/$\textrm{\AA}$, respectively. For PaiNN models this comes with a considerable loss in computational efficiency of the models. For the best PaiNN models, force evaluation times reach even 1000~ms. By choosing lighter settings, PaiNN models with high computational efficiency can be trained, however, the respective RMSEs are not as low as with MACE and REANN models. All-atom force RMSEs obtained with ACE models are only slightly larger than RMSEs obtained with other codes (roughly 27~meV/$\textrm{\AA}$ for the final model), however, H-atom-only force RMSEs are significantly larger (roughly 100~meV/$\textrm{\AA}$) than the RMSEs obtained with NNs for the entire range of ACE settings. ACE models offer the lowest force evaluation times (roughly 4~ms), however, the evaluation times of forces in ACE are higher than those of energies by roughly two times and thus the difference between the computational efficiency of ACE and the most competitive NN models (for which force evaluation times are at roughly 17~ms) is smaller when it comes to force evaluation. Force evaluation times obtained with REANN-Fort are comparable with ACE models (roughly 10~ms), however, REANN-Fort provides accuracy that exceeds the ACE models by a factor of 2. The highest deviations between the force RMSEs obtained for 5 cross-validation models were observed for ACE and REANN, with ACE standard deviations reaching 30~meV/$\textrm{\AA}$ and REANN, even beyond 50~meV/$\textrm{\AA}$. Deep NNs (MACE and PaiNN) are less sensitive to the composition of the training dataset and can achieve more consistent predictions than simpler NN architectures or linear models. The best models in terms of both force accuracy and evaluation speed on CPUs can be listed in the following order (starting with the most successful): REANN-Fort, REANN-Py, MACE, PaiNN, and ACE.

        As mentioned above, utilising high throughput (``task farming'') on very many CPU cores is particularly efficient for the calculation of sticking probability. In studies of larger systems, e.g. containing extended metal slabs, GPUs could be useful. In Supplementary Figure~S11, we plot the energy and force RMSEs against the CPU evaluation times of the final models and we compare their efficiency with the MACE-JAX model evaluated using GPUs. The evaluation times achieved with the MACE-JAX on a GPU can reduce the evaluation times to as low as 1.1~ms, a factor of 50 improvement compared with the CPU case.

For each method, we have selected the most accurate final model that achieves competitive evaluation times of roughly or below 100~ms (marked in Fig.~\ref{fig:error_vs_time} in red color). In the following, we will study the accuracy of these models for a test set and in predicting sticking probabilities.

    \subsection{Accuracy of the optimized models} \label{sec:results_errors}
        \begin{table*}
           \caption{Averaged test RMSEs and MAEs of energies (meV) and forces (meV/$\textrm{\AA}$) using the final PaiNN, REANN, MACE, and ACE models (shown in black in Fig.~\ref{fig:error_vs_time}). The energy table shows the errors for the structures containing H\textsubscript{2} and one of the four surfaces: Cu(111), (100), (110), or (211). The force table shows errors for force predictions per atom for copper atoms of each surface slab (Cu(111), (100), (110), (211)) and separately for only the hydrogen (H) atoms. The last row in both energy and force tables includes combined errors for all the structures (bold font). The errors are averaged over 5 models with different, randomly selected training and test sets (random 3381 and 423 out of 4230 structures, respectively). Standard deviations between the 5 models are shown in parentheses.}
           \begin{tabular}{l|cc|cc|cc|cc} \hline \hline
               \multicolumn{1}{c}{} & \multicolumn{2}{c}{PaiNN} & \multicolumn{2}{c}{REANN} & \multicolumn{2}{c}{ACE} & \multicolumn{2}{c}{MACE} \\ 
               \multicolumn{1}{c}{ } & \multicolumn{2}{c}{Energy}  & \multicolumn{2}{c}{Energy} & \multicolumn{2}{c}{Energy} & \multicolumn{2}{c}{Energy} \\
               Species & RMSE & MAE & RMSE & MAE & RMSE & MAE & RMSE & MAE \\\hline
               H\textsubscript{2}/Cu(111) & 34.4 (24.9) & 27.0 (26.0)  & 15.1 (5.6) & 12.4 (5.9)  & 17.9 (1.3) & 13.1 (1.0)  & 16.6 (4.1) & 11.7 (2.3)  \\
               H\textsubscript{2}/Cu(100) & 29.4 (25.2) & 25.8 (25.8) & 16.0 (9.0) & 10.0 (3.7)  & 18.5 (1.8) & 13.6 (0.7) & 17.3 (3.4) & 11.8 (2.5)   \\
               H\textsubscript{2}/Cu(110) & 33.0 (20.5) & 27.1 (22.5)  &25.5 (11.1) & 17.7 (3.7)  & 29.9 (5.4) & 18.2 (3.0)  & 25.0 (2.5) & 16.3 (1.7)  \\
               H\textsubscript{2}/Cu(211) & 30.8 (21.9) & 25.9 (23.6) & 25.5 (13.4) & 16.6 (5.9)  & 19.6 (4.2) & 13.6 (1.1) & 19.1 (1.0) & 14.1 (1.9)   \\ 
               \textbf{All} & \textbf{31.4 (23.2)} & \textbf{26.5 (24.8)}  & \textbf{18.7 (6.3)} & \textbf{12.3 (3.6)}  & \textbf{18.8 (1.1)} & \textbf{12.0 (0.2)} & \textbf{17.2 (1.7)} & \textbf{11.5 (1.3)} \\ \hline \hline
               \multicolumn{1}{c}{} & \multicolumn{2}{c}{PaiNN} & \multicolumn{2}{c}{REANN} & \multicolumn{2}{c}{ACE} & \multicolumn{2}{c}{MACE} \\ 
               \multicolumn{1}{c}{ } & \multicolumn{2}{c}{Forces}  & \multicolumn{2}{c}{Forces} & \multicolumn{2}{c}{Forces} & \multicolumn{2}{c}{Forces} \\
               Species & RMSE & MAE & RMSE & MAE & RMSE & MAE & RMSE & MAE \\\hline
               Cu(111) & 10.4 (0.8) & 7.6 (0.5) & 7.2 (0.9) & 5.4 (0.7) & 10.6 (0.6) & 7.8 (0.4) & 10.4 (1.0) & 7.7 (0.7) \\
               Cu(100) & 13.0 (1.0) & 9.3 (0.7) & 8.9 (0.5) & 6.4 (0.4) & 12.1 (0.6) & 8.9 (0.5) & 10.9 (0.9) & 8.0 (0.6) \\
               Cu(110) & 14.1 (0.7) & 10.2 (0.5) & 9.0 (0.7) & 6.8 (0.6) & 13.5 (0.5) & 10.2 (0.3) & 10.8 (0.8) & 8.1 (0.5) \\
               Cu(211) & 14.4 (1.5) & 10.6 (1.1) & 9.4 (0.6) & 6.9 (0.5) & 12.5 (0.6) & 9.3 (0.3) & 10.9 (0.7) & 8.2 (0.6) \\ 
               H & 54.8 (29.9) & 19.8 (2.8) & 27.7 (25.4) & 10.3 (1.9) & 91.2 (6.2) & 55.8 (3.9) & 47.0 (34.9) & 20.1 (2.9) \\ 
               \textbf{All} & \textbf{20.3 (5.7)} & \textbf{10.1 (0.5)} & \textbf{11.8 (5.1)} & \textbf{6.5 (0.6)} & \textbf{25.9 (1.5)} & \textbf{12.7 (0.6)} &  \textbf{15.1 (4.2)} & \textbf{8.6 (0.4)} \\ \hline \hline
           \end{tabular}
           \label{tab:error_surfaces}
        \end{table*}
        
        Table~\ref{tab:error_surfaces} reports RMSEs and MAEs of predicted energies and forces with all the methods, averaged over 5 models trained with a random training-validation-test split with corresponding standard deviations of the errors. Different rows in this table report errors for subsets of structures that correspond to different surface facets or different elements, with total averages shown in the last rows and highlighted in bold font. The data shown in these tables corresponds to the models optimized for a trade-off of inference performance, accuracy, and ability to predict the final dynamical observables (sticking probability).
        
        All of the methods achieve energy prediction errors below ``chemical accuracy'' of 1~kcal/mol (43~meV per molecule). The methods ranked from lowest to highest energy prediction error are MACE, ACE, REANN, and PaiNN. The highest energy errors (both RMSEs and MAEs at roughly 30~meV) were obtained with the PaiNN models.  Energy errors achieved with MACE, REANN, and ACE models are comparable (RMSEs within the range of 17-19~meV and MAEs of roughly 12~meV for all), and the highest RMSEs obtained with all of these models belong to H\textsubscript{2}/Cu(110) species and the lowest to H\textsubscript{2}/Cu(110) and H\textsubscript{2}/Cu(100). The highest standard deviation of error predictions was obtained with the PaiNN models (RMSE deviations between 20-25~meV), but consistent across all the surfaces. Significantly lower standard deviations of RMSE predictions were obtained with REANN models, ranging between 5-14~meV. Both MACE and ACE exhibit the most stable energy prediction behavior across all of the surfaces with a standard deviation of RMSE predictions of roughly 1-5~meV.

        Force errors listed in Tab.~\ref{tab:error_surfaces} relate to forces of the individual atoms of different Cu facets or hydrogen atoms. The contribution of hydrogen atoms to the force errors is by far the most significant, due to the difficulty of modeling the barrier and the smaller number of H atoms in every structure (only 2 H atoms per unit cell containing a total of 56 atoms). The different methods demonstrate varying levels of success in the ability to model the hydrogen atom forces. The most successful in predicting hydrogen forces are REANN models, with RMSEs and MAEs only roughly 2 and 3 times, respectively, higher than the surface atom force RMSEs and MAEs (RMSEs 7-9~meV/$\textrm{\AA}$ and MAEs 5-7~meV/$\textrm{\AA}$). MACE and PaiNN predict RMSEs and MAEs for hydrogen forces that are approximately 2 and 5 times higher than the force RMSEs and MAEs of surface atoms for which MACE generates slightly lower RMSEs (11~meV/$\textrm{\AA}$) and MAEs (8~meV/$\textrm{\AA}$) than PaiNN (with RMSEs between 10-14~meV/$\textrm{\AA}$ and MAEs between 8-11~meV/$\textrm{\AA}$). ACE models display fair overall force RMSEs and MAEs (RMSE of 25.9~meV/$\textrm{\AA}$ and MAE of 12.7~meV/$\textrm{\AA}$), however, hydrogen-atom force prediction errors are significantly larger (RMSE and MAE roughly 7 and 6 times larger than corresponding Cu errors). As we will later see, this error will lead to artifacts along the minimum energy path and errors in the dynamics. The origin of such behavior likely lies in the dominance of copper forces in the dataset and the linear fitting procedure. Manual adjustment to increase hydrogen training weights during the least squares fitting procedure of ACE models did not achieve significantly lower hydrogen atom force errors. All of the codes achieve significantly lower errors in comparison with the SchNet MLIPs reported in \cite{stark_machine_2023}.

        To examine the performance of the models in predicting dissociative adsorption barriers for all facets, we evaluated the minimum energy paths (MEPs), using climbing image nudged elastic bands (CI-NEBs) with all the considered models and compared the results with the DFT reference (Supplementary Figure~S12). The MEPs predicted with PaiNN, MACE, and REANN agree with the DFT reference MEPs very well for all surface facets, confirming that chemical accuracy along the dissociation path has been obtained with these models. The ACE model provides an accurate prediction of MEPs for hydrogen adsorption on Cu(111) and Cu(211) surfaces, however, significant deviations in predicting MEPs for hydrogen adsorption on Cu(100) and Cu(110) arise. The MEP corresponding to dissociation on the Cu(100) surface is still fairly accurate, giving the correct activation energy, however, the barrier is shifted towards the product. The MEP generated with the ACE model for hydrogen dissociation on the Cu(110) surface features a physisorption artifact at around 2.8~$\textrm{\AA}$ and the reaction barrier is underestimated by roughly 0.17~eV, which may cause significant difficulties in predicting sticking probabilities which are highly sensitive to the dissociation barrier.
        
        \begin{figure*}
            \centering
            \includegraphics[width=1.0\linewidth]{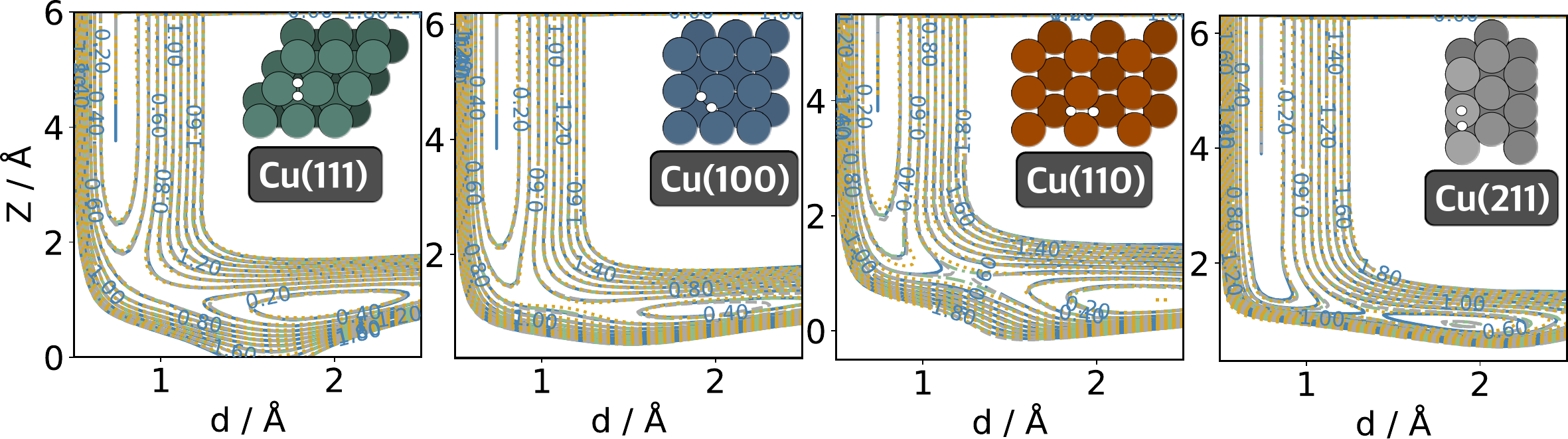}
                \caption{\textbf{Elbow plots representing energy landscape for dissociative adsorption of H\textsubscript{2} on different Cu facets.} The energy levels are shown within the distance between hydrogen atoms (d) and the distance between the center of mass of the hydrogen molecule from the surface. Contour lines represent results obtained with ACE (yellow, dotted line), MACE (green, dash-dotted line), REANN (gray, dashed line), and PaiNN (blue, solid line). Labels showing energy values (in eV) on the contour lines correspond to the PaiNN lines.}
            \label{fig:elbow_plots}
        \end{figure*}

        To further assess the accuracy of the trained models to predict the PESs around the transition state region, Figure~\ref{fig:elbow_plots} presents two-dimensional cuts through the PES along the center-of-mass distance of the molecule from the surface, Z, and the intramolecular distance, d, (so-called ``elbow plots'') for all the considered Cu facets. The PESs generated by all the models are smooth, unlike the corresponding PESs predicted by SchNet models trained on the same data~\cite{stark_machine_2023}. The PES contours generated with different codes match each other closely, with just a few exceptions around the reaction barrier on Cu(110) and Cu(211) surfaces and close to the product states on the Cu(211) surface. In both cases, the differences are not significant enough to conclude if they determine the final predictions of sticking probabilities. We can, however, compare the PESs for the Cu(110) with the corresponding ACE-generated MEP (Supplementary Figure~S12), which fails to accurately follow the DFT-generated MEPs. In this case, we can see that the energy contours preceding the reaction barrier are slightly shifted towards the barrier and may contain artificial minima, as indicated in the corresponding MEP. 
        Therefore, ACE models will most likely require additional training data to correctly predict the Cu(110) barrier region and dynamical observables that sensitively depend on it.

        Finally, we predict phonon bandstructures with the ACE, MACE, and REANN MLIPs and compare them against the DFT reference (Supplementary Figure~S13). The overall ability of the models to predict the frequencies and band dispersion is excellent for all four surfaces with only some minor discrepancies visible. MACE and REANN predictions are almost identical. ACE model predictions slightly diverge from the predictions made with the other methods, which is visible, particularly for Cu(100) and Cu(110) surfaces. Nevertheless, even ACE can reproduce phonon bandstructures very accurately. This shows that all MLIPs faithfully capture the temperature-dependent structure and phonon dynamics of the metal substrates and only differ in their ability to capture the dissociative chemisorption dynamics of hydrogen.
        

    \subsection{Comparison with previous theoretical results}  \label{sec:results_stick_theor}

        \begin{figure}
            \centering
            \includegraphics[width=0.6\linewidth]{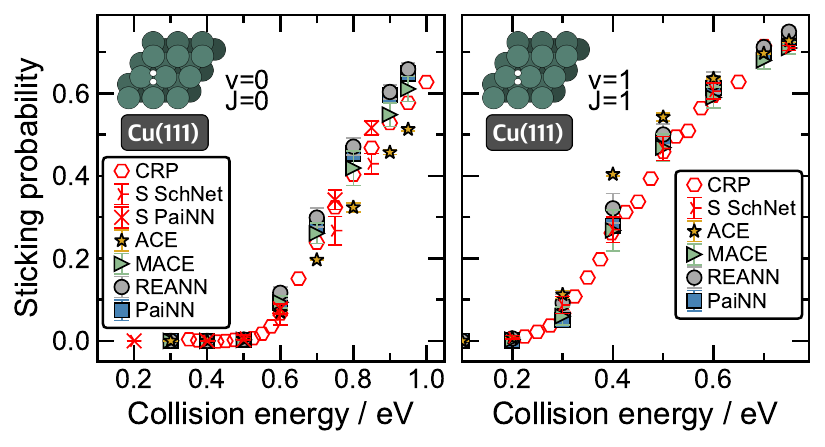}
            \caption{\textbf{Sticking probabilities for H\textsubscript{2} scattering on Cu(111) at 0~K.} Probabilities were calculated at different collision energies using ACE (yellow stars), MACE (green triangles), REANN (gray circles), and PaiNN (blue squares) models for the ground ($\mathrm{\nu}$=0 and J=0) (left) and excited ($\mathrm{\nu}$=1 and J=1) (right) rovibrational initial states. Each point represents an averaged value of sticking probability predicted using a committee of 3 models based on different train/test splits. Error bars correspond to standard deviations between the predictions made with all the models. For each initial condition, 10,000 trajectories were used for ensemble averaging. QCD-BOSS reference simulation results obtained by Smits~\textit{et~al.}~\cite{smits_quantum_2022} are depicted as red hexagons (CRP). PaiNN (S PaiNN) and SchNet (S SchNet) reference data from Ref.~\cite{stark_machine_2023} are depicted as red crosses and tri-crosses, respectively.}
            \label{fig:sticking_cu111_t0}
        \end{figure}

        Having established the accuracy of the MLIPs to predict energy landscapes and forces for an in-distribution test dataset, in Fig.~\ref{fig:sticking_cu111_t0}, we study their ability to predict H\textsubscript{2} sticking probabilities at 0~K Cu(111) surfaces for 8 different translational energies and two rovibrational initial states, ($\mathrm{\nu}$=0 and J=0) and ($\mathrm{\nu}$=1 and J=1). Therefore, we can compare to previous results obtained with PaiNN and SchNet MLIPs from Ref.~\cite{stark_machine_2023} and with the results obtained for a CRP with Born-Oppenheimer static surface (BOSS) approximation, reported by Smits~\textit{et~al.}~\cite{smits_quantum_2022}. Predictions of sticking probabilities for MACE, REANN, and PaiNN models are in excellent agreement for both rovibrational states, with MACE being the closest. REANN and PaiNN predict slightly higher sticking probabilities than the reference results by Smits~\textit{et~al.}~\cite{smits_quantum_2022} and PaiNN results by Stark~\textit{et~al.}~\cite{stark_machine_2023}. The ACE model slightly underestimates sticking probabilities for the ground rovibrational state ($\mathrm{\nu}$=0 and J=0), especially for higher collision energies, above 0.6~eV. The barrier of hydrogen dissociation on Cu(111) is 0.64~eV (bridge site)~\cite{smeets_designing_2021}. We can thus see the high correlation between the ability to model barriers accurately and the accuracy in predicting sticking probability. For H\textsubscript{2}($\mathrm{\nu}$=1 and J=1), ACE models overestimate sticking probabilities considerably, mainly for the moderate collision energies (e.g. 10\% higher sticking probability at 0.4 eV collision energy).

    \subsection{Performance of the models in comparison with high temperature experiments}  \label{sec:results_stick_exp}

        \begin{figure}
            \centering
            \includegraphics[width=0.6\linewidth]{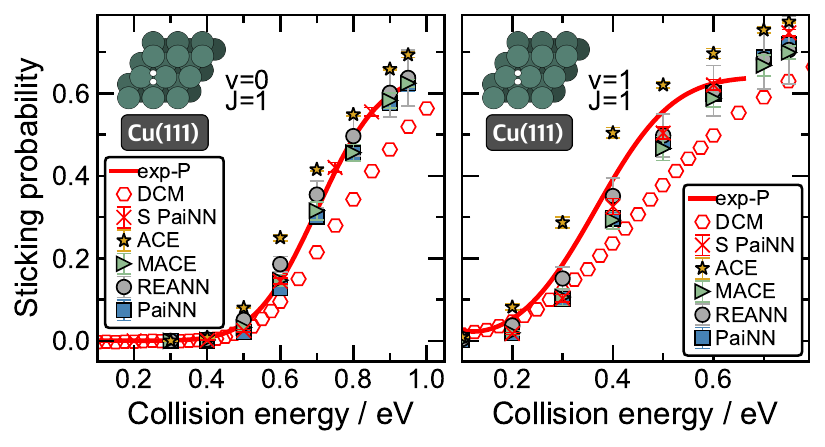}
            \caption{\textbf{Sticking probabilities for H\textsubscript{2} scattering on Cu(111) at 925~K.} Probabilities were calculated at different collision energies using ACE (yellow stars), MACE (green triangles), REANN (gray circles), and PaiNN (blue squares) models for the ground ($\mathrm{\nu}$=0) (left) and excited ($\mathrm{\nu}$=1) (right) vibrational states (J=1 in both cases). Each point represents an averaged value of sticking probability predicted using a committee of 3 models based on different train/test splits. Error bars correspond to standard deviations between the predictions made with all the models. For each initial condition, 10,000 trajectories were used for ensemble averaging.
            The red line represents a sticking function obtained from the experimental results of  Kaufmann~\textit{et~al.}~\cite{kaufmann_associative_2018} (exp-P) at 923$\pm$3~K, scaled to match the ``S PaiNN'' sticking probabilities at the highest incidence energy (saturation parameter A=0.64 for both sticking functions). DCM reference results obtained by Smits~\textit{et~al.}~\cite{smits_quantum_2022} are depicted as red hexagons. PaiNN (``S PaiNN'') reference data from Ref.~\cite{stark_machine_2023} is depicted as red crosses.}
            \label{fig:sticking_cu111_exp}
        \end{figure}

\begin{figure}
            \centering
            \includegraphics[width=0.35\linewidth]{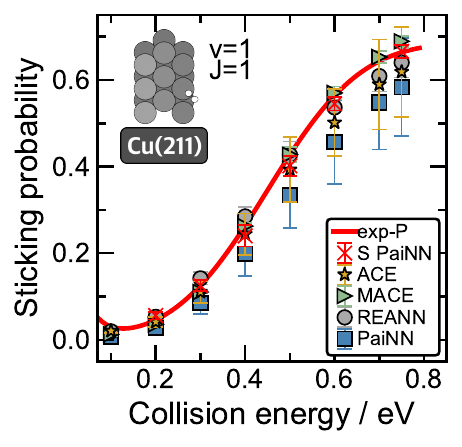}
            \caption{\textbf{Sticking probabilities for H\textsubscript{2} scattering on Cu(211) at 925~K.} Probabilities were calculated at different collision energies using ACE (yellow stars), MACE (green triangles), REANN (gray circles), and PaiNN (blue squares) models for the ($\mathrm{\nu}$=1, J=1) rovibrational states. Each point represents an averaged value of sticking probability predicted using a committee of 3 models based on different train/test splits. For each initial condition, 10,000 trajectories were used for ensemble averaging. Error bars correspond to standard deviations between the predictions made with all the models. The red line represents a sticking function obtained from the experimental results reported by Kaufmann~\textit{et~al.}~\cite{kaufmann_associative_2018} (exp-P) at 923$\pm$3~K, scaled to match the theoretical probabilities from~\cite{stark_machine_2023}, specifically, the ``S PaiNN'' results at the highest incidence energy (saturation parameter A=0.66 for both sticking functions). PaiNN (``S PaiNN'') reference data from Ref.~\cite{stark_machine_2023} is depicted as red crosses.}
            \label{fig:sticking_v1j1_cu211}
        \end{figure}

        To independently assess the MLIPs against experimental data, we simulated sticking probabilities of H\textsubscript{2} on Cu(111) in the ground and first vibrational excited states (see Figure~\ref{fig:sticking_cu111_exp}) and for H\textsubscript{2} on Cu(211) in the first excited vibrational state (Figure~\ref{fig:sticking_v1j1_cu211}).  We compare our new results with the results obtained with PaiNN models from Ref.~\cite{stark_machine_2023}, and a DCM model based on a CRP potential for hydrogen atoms and an embedded atom method (EAM) for the surface degrees of freedom, reported by Smits~\textit{et~al.}~\cite{smits_quantum_2022}.
        
        It is worth noting that the sticking probabilities obtained from permeation experiments cannot be directly compared to simulations.~\cite{kaufmann_associative_2018} The common approach for comparing results with permeation experiments is to scale the experimental results by adjusting the saturation parameter of the experimental sticking curve based on the highest reported collision energy.~\cite{kaufmann_associative_2018,zhu_unified_2020,stark_machine_2023} Since the accuracy of PaiNN models from our previous study is already established, we have scaled the experimental results to the sticking probabilities calculated with PaiNN (``S PaiNN'') for the highest collision energy. We note that the previous PaiNN results did not explicitly include lattice expansion at 925~K, whereas the new PaiNN results do. It was previously noted by Mondal~\textit{et~al.}~\cite{mondal_thermal_2013} that lattice expansion can significantly affect sticking probabilities, however, in this case, the effect on the predictions is negligible.
        
        
        The sticking probabilities on Cu(111) at 925~K evaluated using the MACE, REANN, and new PaiNN models match the reference experimental curve and previous PaiNN results closely for both the ground and first excited vibrational states. ACE models more considerably overestimate sticking probabilities at the ground and first excited vibrational states than in the 0~K case. The highest deviation from the reference results is at the collision energy of 0.6~eV for the ground vibrational state and of 0.3~eV for the first excited state, with both overestimating the probability by roughly 0.12~eV. Only REANN models cause significant uncertainties for the predictions, which was also noticeable during cross-validation (Fig.~\ref{fig:error_vs_time}). The uncertainties, however, are not large enough to significantly affect the outcome of the sticking probability evaluation or to diminish the predictive power of the simulations.

        For H\textsubscript{2} scattering on Cu(211) at 925~K  in a rovibrationally excited state ($\mathrm{\nu}$=1, J=1) (Fig.~\ref{fig:sticking_v1j1_cu211}), MACE and REANN models once again provide excellent agreement with the reference. The ACE predictions qualitatively follow the sticking curve when compared to the reference methods, however with considerable prediction uncertainty of around $\pm$10\%. The newly trained, lighter PaiNN models underestimate the sticking probabilities by roughly 10\% between the collision energies of 0.5 and 0.7~eV. Additionally, lighter PaiNN models cause significant uncertainties in sticking probability, of roughly 0.1. Therefore, using the lighter PaiNN models may not be sufficient to model all the studied surfaces as accurately as with REANN or MACE models, which further increases time-to-solution for simulations with accurate PaiNN models. The performance of REANN and MACE models is excellent in terms of the agreement with the experiment at much higher computational efficiency than both larger and smaller PaiNN models (``S PaiNN'' and ``PaiNN'', respectively).


\section{\label{sec:conclusions}Conclusions}

Recent advances in machine learning interatomic potentials (MLIPs), such as message passing~\cite{gilmer_neural_2017,schutt_quantum-chemical_2017}, equivariant features~\cite{batzner_e3-equivariant_2022,schutt_equivariant_2021} or full-body descriptors (ACE)~\cite{drautz_atomic_2019} allowed significant improvement in both accuracy and computational efficiency of MLIPs. In this study, we explored some of the most commonly used MLIPs and their ability to describe reactive chemical dynamics at surfaces. This provides an interesting challenge for MLIPs as the simulation of nonequilibrium scattering probabilities at surfaces requires comprehensive statistical sampling for many different initial conditions, which easily amounts to millions of molecular dynamics simulations of large condensed phase models. As reaction probabilities are highly sensitive to subtle barriers, suitable MLIPs need to be chemically accurate yet provide highly efficient energy and force evaluation in the low millisecond regime.

Based on an existing training dataset for reactive scattering of molecular hydrogen on four low-index facets of copper, we studied the ability of the PaiNN,  MACE, and REANN message-passing neural networks, and the Atomic Cluster Expansion (ACE) potentials to deliver accurate and highly efficient MLIPs for gas-surface dynamics. 

All models were able to capture the structure and dynamics of the metal surfaces over a wide temperature range, predicting lattice expansion and phonon bandstructures with high accuracy. The hydrogen dynamics along the reaction path provided a greater challenge for the models. We show that through careful hyperparameter optimization, an optimal trade-off between the accuracy and inference performance of MLIPs can be identified. In the cases of MACE, REANN, and PaiNN, we were able to train models that provide low in-distribution errors and accurate dynamical predictions of reactive sticking probabilities when compared to previous results from simulation and experiment. However, MACE and REANN achieve a given accuracy with consistently lower time-to-solution than PaiNN, namely about 0.5 and 0.2~ms, respectively, per force evaluation per atom for the most efficient models when evaluated on CPUs. MACE achieves this with graph-based message passing and an efficient atom-centered many-body representation. REANN on the other hand uses a more compact network layout. Both are viable options and should be considered state-of-the-art to construct MLIPs for reactive chemical dynamics simulations. Potentials constructed via the Atomic Cluster Expansion within ACEPotentials.jl can potentially deliver even faster energy and force evaluations (ca. 0.1~ms per force evaluation per atom when evaluated on CPUs). In the present study, it did not yield a consistently accurate representation of the PES along relevant reaction coordinates when trained on the same training data as the other models and may require additional training data or further hyperparameter optimisation to reach the accuracy of NN-based methods. 

The employed training dataset was generated with active learning based on an MPNN MLIP. A dataset generated directly through active learning with ACE might have yielded better prediction results. While MACE and REANN provide sufficiently robust and efficient solutions for many problems in gas-surface dynamics, for some extremely demanding cases (large system sizes or demanding statistical sampling requirements), curation of balanced training datasets to construct ACE potentials may be a good alternative to achieve even lower time-to-solution.


\section*{Acknowledgments}
    This work was financially supported by The Leverhulme Trust (RPG-2019-078), the UKRI Future Leaders Fellowship programme (MR/S016023/1 and MR/X023109/1), and a UKRI frontier research grant (EP/X014088/1). High-performance computing resources were provided via the Scientific Computing Research Technology Platform of the University of Warwick, the EPSRC-funded Materials Chemistry Consortium (EP/R029431/1, EP/X035859/1) and the UK Car-Parrinello consortium (EP/X035891/1) for the ARCHER2 UK National Supercomputing Service , and the EPSRC-funded HPC Midlands+ computing centre for access to Sulis (EP/P020232/1). We thank Matthias Sachs (Birmingham) and Christoph Ortner (UBC, Vancouver) for helpful discussions regarding the ACE models.

\section*{Data and Code Availability}
The molecular dynamics simulations were performed with the publicly available open-source code NQCDynamics.jl. All employed data was previously published. The source code and documentation are available on Github: \url{https://nqcd.github.io/NQCDynamics.jl/stable/}.

\section*{ORCID iDs}
Wojciech G. Stark \url{https://orcid.org/0000-0001-6279-2638}\\
Cas van der Oord \url{https://orcid.org/0000-0003-1845-0387}\\
Ilyes Batatia \url{https://orcid.org/0000-0001-6915-9851}\\
Yaolong Zhang \url{https://orcid.org/0000-0002-4601-0461}\\
Bin Jiang \url{https://orcid.org/0000-0003-2696-5436}\\
Gábor Csányi \url{https://orcid.org/0000-0002-8180-2034}\\
Reinhard J. Maurer \url{https://orcid.org/0000-0002-3004-785X}

\section*{Author Contributions}
\textbf{Wojciech G. Stark:}
conceptualization (equal);
data curation (lead);
investigation (lead);
software (lead);
validation (lead);
visualization (lead);
writing -- original draft (lead);
writing -- review and editing (equal).
\textbf{Cas van der Oord:} 
investigation (supporting);
software (supporting).
\textbf{Ilyes Batatia:}
investigation (supporting);
software (supporting);
writing -- review and editing (supporting).
\textbf{Yaolong Zhang:}
investigation (supporting);
software (supporting);
writing -- review and editing (supporting).
\textbf{Bin Jiang:}
writing -- review and editing (supporting).
\textbf{Gábor Csányi:}
writing -- review and editing (supporting).
\textbf{Reinhard J. Maurer:}
conceptualization (equal);
investigation (equal);
supervision (lead);
writing -- review and editing (equal).

\section{Conflicts of Interest}
CvdO and GC have controlling interest in Symmetric Group LLP that markets ML force fields. 

\def\bibsection{\section*{\refname}} 
\bibliography{main}


\end{document}


\title[Supplementary information]{Supplementary Information to \\"Benchmarking of machine learning interatomic potentials for reactive hydrogen dynamics at metal surfaces"}

\author{Wojciech G. Stark}%
\address{Department of Chemistry, University of Warwick, Gibbet Hill Road, Coventry CV4 7AL, United Kingdom}

\author{Cas van der Oord}%
\address{Department of Engineering, University of Cambridge, United Kingdom}

\author{Ilyes Batatia}%
\address{Department of Engineering, University of Cambridge, United Kingdom}

\author{Yaolong Zhang}%
\address{Department of Chemistry and Chemical Biology, Center for Computational Chemistry, University of New Mexico, Albuquerque, New Mexico 87131, USA}

\author{Bin Jiang}%
\address{Key Laboratory of Precision and Intelligent Chemistry, Department of Chemical Physics, University of Science and Technology of China, Hefei, Anhui, China}
\address{Hefei National Laboratory, University of Science and Technology of China, Hefei 230088, China}

\author{Gábor Csányi}%
\address{Department of Engineering, University of Cambridge, United Kingdom}
    
\author{Reinhard J. Maurer}%
\email{r.maurer@warwick.ac.uk}
\address{Department of Chemistry, University of Warwick, Gibbet Hill Road, Coventry CV4 7AL, United Kingdom}
\address{Department of Physics, University of Warwick, Gibbet Hill Road, Coventry CV4 7AL, United Kingdom}

\date{\today}
\maketitle


\section{Impact of non-zero spin structures} \label{sec:si_outliers}
    \begin{figure}
        \centering
        \includegraphics[width=1.0\linewidth]{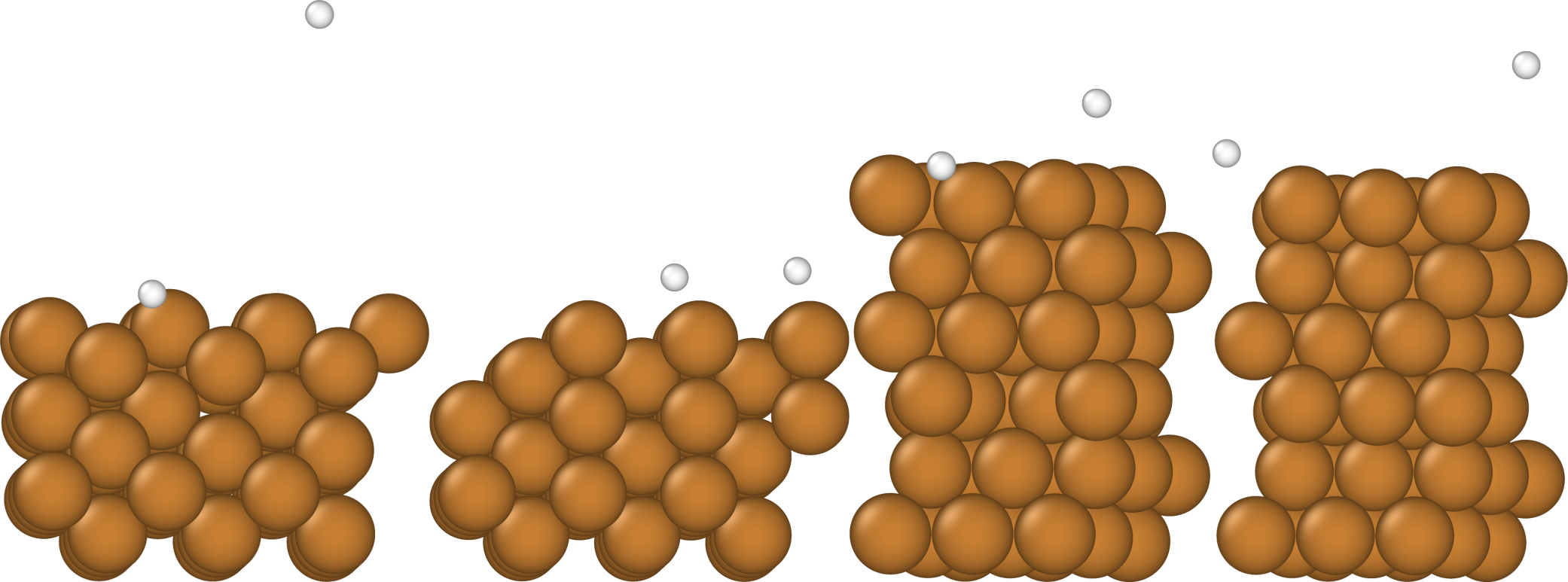}
            \caption{\textbf{Examples of non-zero spin structures from our data set.} Four out of ten non-zero spin structures from our data set are shown, containing two H atoms at Cu(110) (first two structures) and Cu(111) surfaces (last two structures).}
        \label{fig:highspin_si}
    \end{figure}
    We identified 10 non-zero spin structures in our data set. Four examples of such structures, containing two hydrogen atoms at Cu(110) (first two structures) and Cu(111) (last two structures) are shown in Fig.~\ref{fig:highspin_si}.

    \begin{figure}
        \centering
        \includegraphics[width=1.0\linewidth]{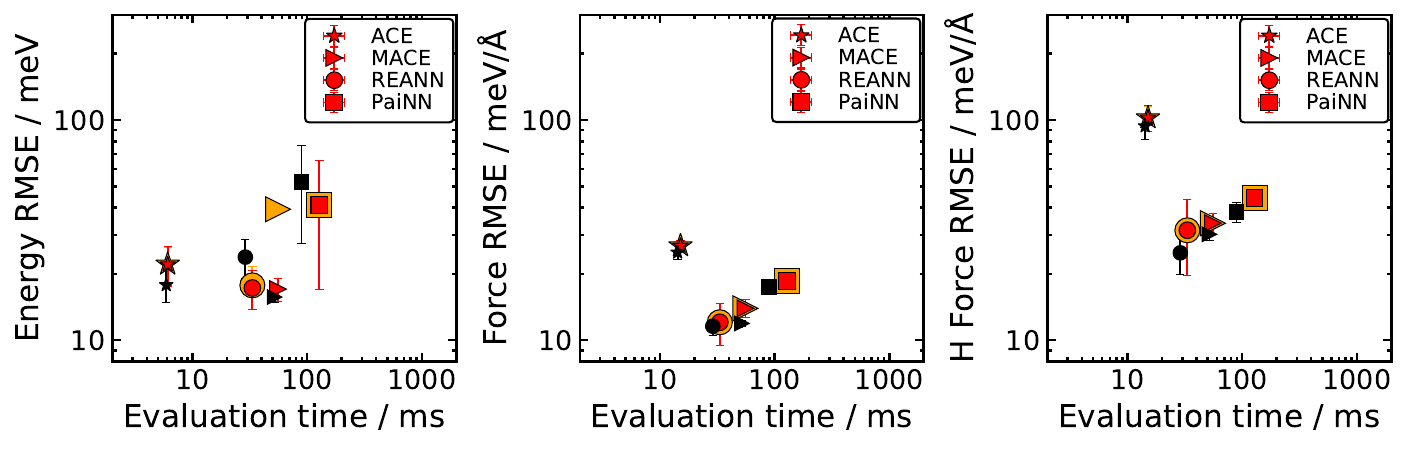}
            \caption{\textbf{Energy and force (all atoms and H only) test RMSEs in meV and meV/$\textrm{\AA}$ plotted against evaluation time (CPU) in milliseconds for models trained with and without high-spin structures.} We include the results obtained with models trained for cross-validation, using PaiNN (squares), REANN (circles), MACE (triangles), and ACE (stars). Data points and error bars correspond to the RMSE average and standard deviation over 5-fold cross-validation splits. The largest, orange-colored markers correspond to the final optimized models, trained with the non-zero spin structures, and including the non-zero spin structures in the test set. Slightly smaller, red-colored markers indicate RMSEs calculated with the same model, but the non-zero spin structures are omitted in the test set evaluation. The smallest, black-colored markers correspond to the results obtained with models trained using the same settings but without the high-spin structures in the data set. The energy and force evaluation times were calculated using a single AMD EPYC 7742 (Rome) 2.25 GHz processor core.}
        \label{fig:error_vs_time_si}
    \end{figure}
    To check the validity of the final ACE, MACE, REANN, and PaiNN models, we plotted energy, hydrogen atom force, and overall force RMSEs with respect to single evaluation times (Fig.~\ref{fig:error_vs_time_si}) calculated in three ways: by a model that includes non-zero spin structures in both training and test sets (orange markers), by the same model (non-zero spin structures in a training set), but for the test set without the non-zero spin structures (red markers), and finally by a model without such structures in both training and test sets (black markers). This allowed us to compare if the non-zero spin structures impact the learning ability of the models. For the models without the high-spin structures, RMSEs are comparable, even though one model was trained with outlier structures and the other was not. 

\newpage
\section{Dependence of energy with respect to the copper lattice constant} \label{sec:si_lattice}

    \begin{figure*}
        \centering
        \includegraphics[width=3.1in]{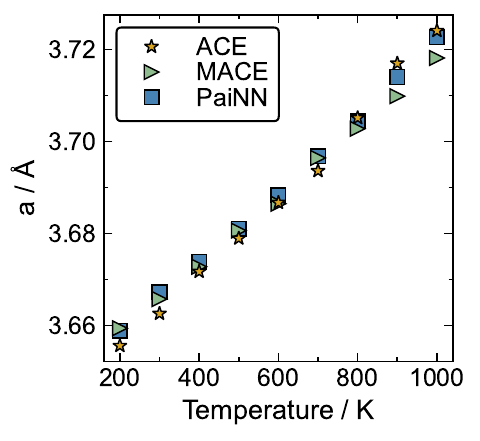}
        \caption{\textbf{Dependence of lattice constant, a ($\textrm{\AA}$), on temperature}. The lattice constants were extracted from NPT simulations using ACE (yellow stars), MACE (green triangles), and PaiNN (blue squares) models in 9 temperatures ranging from 200 and 1000~K.}
        \label{fig:lattice_temp}
    \end{figure*}

    To obtain the lattice constants for the MD simulations at higher temperatures, we plotted the lattice constants at different temperatures obtained from NPT simulations using ACE, MACE, and PaiNN models (Fig.~\ref{fig:lattice_temp}). For the NPT simulations, we used a time step of 1~fs and a barostat parameter of 2E6~GPa$\cdot$fs$^{2}$. The simulations were run for 20~ps to ensure thermal equilibration. The agreement between the predictions using different models is very good and the final lattice constant for the MD simulations predicted by all the models in 925~K is 3.71~$\textrm{\AA}$. 

    \begin{figure*}
        \centering
        \includegraphics[width=3.1in]{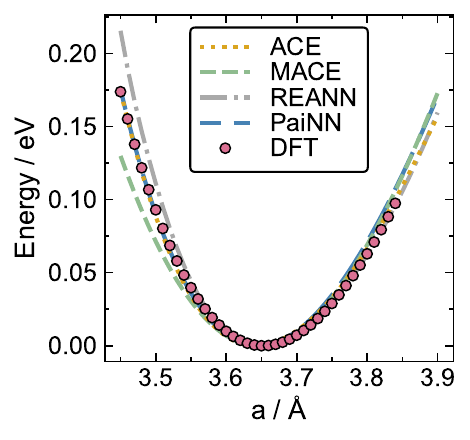}
        \caption{\textbf{Relative potential energies of the Cu atoms for a set of lattice constants, a ($\textrm{\AA}$)}. Energies were calculated using DFT (red circles) and all the MLIPs included in our study, namely, ACE (yellow, dotted line), MACE (green, shortly-dashed line), REANN (gray, "dash-dot" line), and PaiNN (blue, dashed line).}
        \label{fig:lattice}
    \end{figure*}

     The relative potential energies evaluated for the range of lattice constant parameters using DFT and ACE, MACE, REANN, and PaiNN models are shown in Fig.~\ref{fig:lattice}. We can clearly observe that energy predictions for different lattice constants are excellent.

\section{Learning behavior and model parameter optimization} \label{sec:si_optimization}

    We performed an optimization of different model parameters with all the codes included in this study, namely PaiNN, MACE, ACE, and REANN. We incorporate three measures for determining the accuracy of the models, energy, force, and hydrogen-atom-only force RMSEs. Additionally, we include a model efficiency measure, which is the energy evaluation time. The convergence tests shown below include the optimization of settings with the greatest impact on the model accuracy. All the evaluation times have been obtained using a single core of the AMD EPYC 7742 (Rome) 2.25 GHz processor.
    
        \begin{figure*}
            \centering
            \includegraphics[width=5.8in]{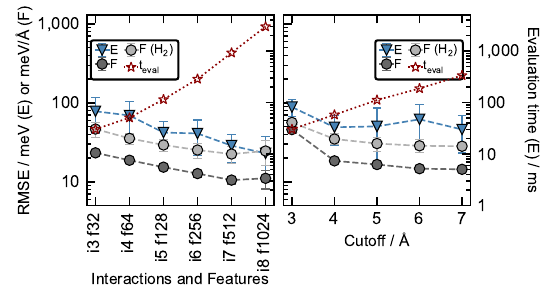}
            \caption{\textbf{Optimization of PaiNN model parameters.} Convergence with respect to the number of interaction layers and features (left) and cutoff distance (right), based on the energy (blue circles), force (dark-gray circles), and hydrogen-only force (light-gray circles) test RMSEs. Additionally, corresponding average energy evaluation times, based on 100 energy evaluations, are included (red stars).}
            \label{fig:opt_painn_1}
        \end{figure*}
        
    \subsection{PaiNN} \label{sec:si_optimization_painn}
    
        The PaiNN model performance depends sensitively on the number of interaction layers and features and cutoff distance. The relation between the energy and force test RMSEs and different model settings was included in Fig.~\ref{fig:opt_painn_1}. On the left side of Fig.~\ref{fig:opt_painn_1}, we plotted the RMSEs obtained using models, which differ in the number of interaction layers and features. The convergence seems relatively slow with convergence at the 7 interaction layers and 512 features, however, decent RMSEs can be achieved even with 4 interaction layers and 64 features. On the right side of Fig.~\ref{fig:opt_painn_1}, we investigate the convergence of cutoff distance, which similarly shows that the convergence is relatively slow and is achieved at 6~$\textrm{\AA}$, however, the final energy and force RMSEs can be achieved at a satisfactory level even using a cutoff of 4~$\textrm{\AA}$.

    \subsection{REANN} \label{sec:si_optimization_reann}
        \begin{figure*}
            \centering
            \includegraphics[width=5.8in]{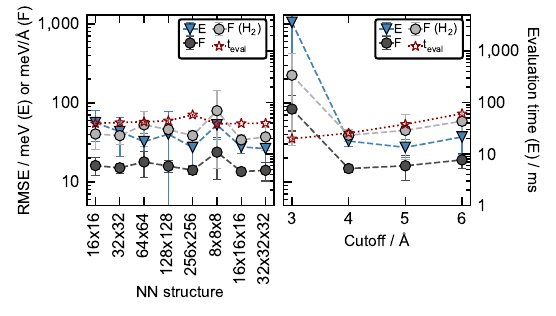}
            \caption{\textbf{Optimization of REANN model parameters.} Convergence of NN structure (left) and cutoff distance (right), based on the energy (blue circles), force (dark-gray circles), and hydrogen-only force (light-gray circles) test RMSEs. Additionally, corresponding average energy evaluation times, based on 100 energy evaluations, are included (red stars).}
            \label{fig:opt_reann_1}
        \end{figure*}
        
\begin{figure*}
            \centering
            \includegraphics[width=5.8in]{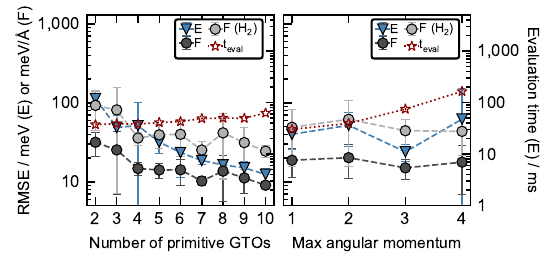}
            \caption{\textbf{Optimization of REANN model parameters.} Convergence of the number of primitive GTOs (left) and maximum angular momentum (L) (right), based on the energy (blue circles), force (dark-gray circles), and hydrogen-only force (light-gray circles) test RMSEs. Additionally, corresponding average energy evaluation times, based on 100 energy evaluations, are included (red stars).}
            \label{fig:opt_reann_2}
        \end{figure*}

        The optimization of REANN model parameters included the convergence tests of the size of the NN structure (number of neurons in each hidden layer), the cutoff distance (Fig.~\ref{fig:opt_reann_1}), the number of primitive Gaussian-type orbitals (GTOs), and maximum angular momentum (L) of primitive GTOs Fig.~\ref{fig:opt_reann_2}. We plot the test force (from all atoms and just hydrogen) and energy RMSEs for different sizes of the NN (left side of Fig.~\ref{fig:opt_reann_1}). Using 3 hidden layers in the NN provides better efficiency for the same accuracy than using 2 hidden layers, More specifically, using a 16$\times$16$\times$16 architecture allows us to achieve the accuracy of the 256$\times$256 with the efficiency of the 16$\times$16. On the right side of Fig.~\ref{fig:opt_reann_1} we plot the test RMSEs for varying cutoff distances. REANN models achieve quick convergence with a cutoff of 4~$\textrm{\AA}$.
        
        The optimization of the number of primitive GTOs (nwaves) is shown on the left side of Fig.~\ref{fig:opt_reann_2}. Convergence of this parameter is achieved relatively slowly and leads to the conclusion that the models require at least 7 primitive GTOs. The maximum angular momentum (L) (right side of Fig.~\ref{fig:opt_reann_2}) turned out not to be an important factor in the final test RMSE and can be used at the value of 1.

    \subsection{ACE} \label{sec:si_optimization_ace}
        \begin{figure*}
            \centering
            \includegraphics[width=5.8in]{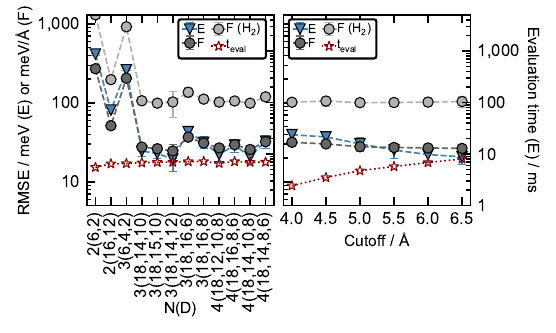}
            \caption{\textbf{Optimization of ACE model parameters.} Convergence of correlation order and maximum polynomial degree (left) and cutoff distance (right), based on the energy (blue circles), force (dark-gray circles), and hydrogen-only force (light-gray circles) test RMSEs. Additionally, corresponding average energy evaluation times, based on 100 energy evaluations, are included (red stars).}
            \label{fig:opt_ace_1}
        \end{figure*}
        
        The optimization of parameters for ACE models included optimizing correlation order, the corresponding maximum polynomial degree, and cutoff distance. On the left side of Fig.~\ref{fig:opt_ace_1} the convergence of both correlation order and maximum polynomial degrees is depicted. The models require a correlation order of at least 3. We noticed that in terms of the efficiency of the models, it is favorable to use descending polynomial degrees instead of the same degrees in each correlation order. Because of the linearity of the models and the negligible difference in the efficiency of the models, we decided to use a correlation order of 4 with the maximum polynomial degrees of 18, 12, 10, and 8. The convergence of cutoff distance (right side of Fig.~\ref{fig:opt_ace_1}) is much slower than for other NN-based models that include message passing in the model training, which is to be expected. We find that the convergence is reached at 5.5-6~$\textrm{\AA}$.
    
    \subsection{MACE} \label{sec:si_optimization_mace}

        \begin{figure*}
            \centering
            \includegraphics[width=5.8in]{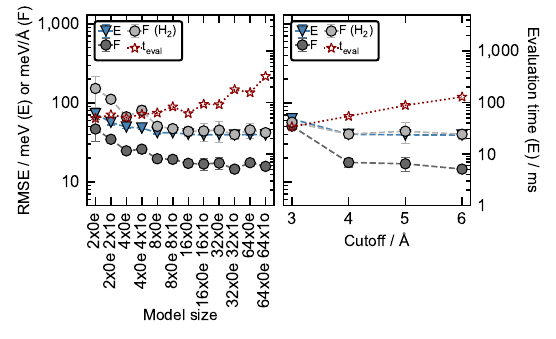}
            \caption{\textbf{Optimization of MACE model parameters.} Convergence of model size (left) and cutoff distance (right), based on the energy (blue circles), force (dark-gray circles), and hydrogen-only force (light-gray circles) test RMSEs. Additionally, corresponding average energy evaluation times, based on 100 energy evaluations, are included (red stars).}
            \label{fig:opt_mace_1}
        \end{figure*}
        
        We show the convergence of the parameters used for training of MACE models in Fig.~\ref{fig:opt_mace_1}, including the model size, including invariant (e.g. \textit{2x0e}) and equivariant messages (e.g. \textit{2x0e 2x1o}) and cutoff distance. On the left side of Fig.~\ref{fig:opt_mace_1} we plotted the energy and force RMSEs, as well as the energy evaluation times for different model sizes. Including equivariant features decreases the RMSEs for smaller model sizes, however, for the larger model sizes the RMSEs obtained with invariant and equivariant models are comparable. On the right side of Fig.~\ref{fig:opt_mace_1}, we plotted the force and energy test RMSEs, as well as the energy evaluation times for varying cutoff distances. Similarly to other message-passing-based codes (PaiNN and REANN), we obtain satisfactory convergence already at 4~$\textrm{\AA}$.

    \subsection{Learning curves based on MAEs} \label{sec:si_optimization_mace}

    \begin{figure}
        \centering
        \includegraphics[width=1.0\linewidth]{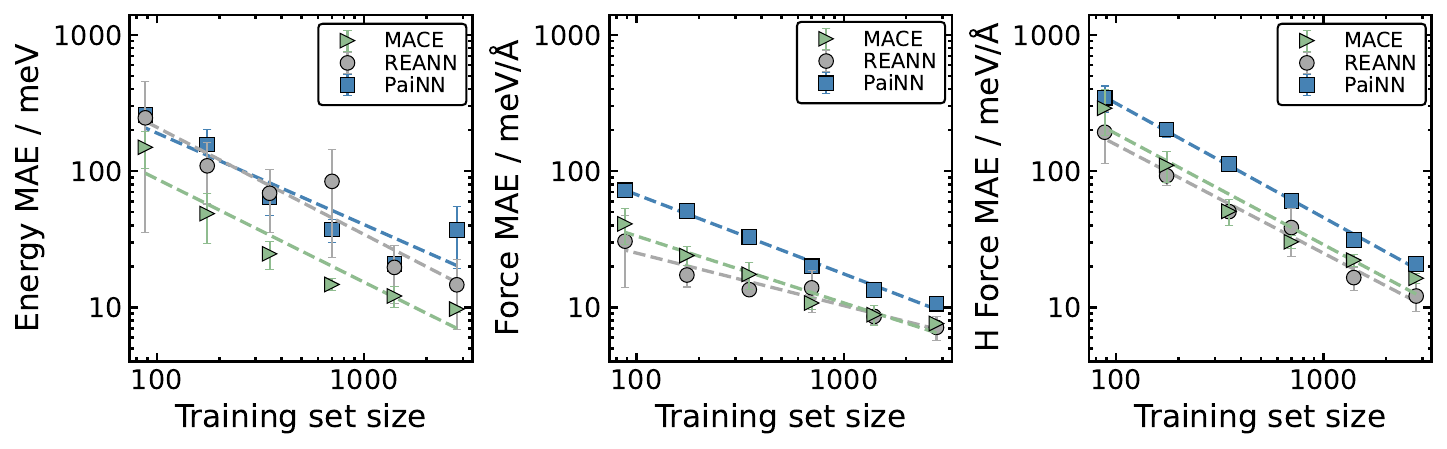}
            \caption{\textbf{Learning curves for MACE, REANN, and PaiNN models.} Log-log plot of the averaged test set predictive error (MAE) in energy, force, and hydrogen-atom-only force as a function of training set size for all NN-based models employed in this study. The shown MAE values are averaged over 5 cross-validation splits, differing by training and validation sets. Error bars correspond to standard deviations between the MAEs obtained over all splits.}
        \label{fig:learning_curves_mae_si}
    \end{figure}

    Learning curves based on MAEs were plotted using the results generated with MACE, REANN, and PaiNN models (Fig.~\ref{fig:learning_curves_mae_si}). The MAE-based learning curves show overall lower errors than RMSE-based learning curves. This is caused by MAEs being less sensitive to outliers than RMSEs. 

\newpage  
\section{Efficiency with GPU architecture} \label{sec:si_gpu}

    \begin{figure}
        \centering
        \includegraphics[width=1.0\linewidth]{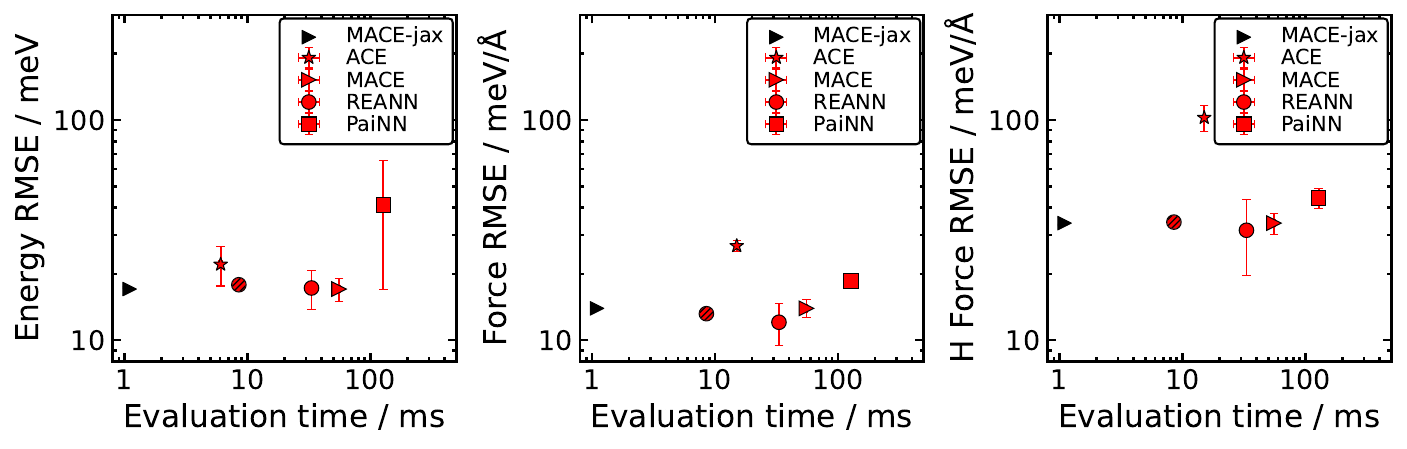}
            \caption{\textbf{Energy and force (all atoms and H only) test RMSEs in meV and meV/$\textrm{\AA}$ plotted against the CPU evaluation time of the final models and GPU evaluation time of the MACE-JAX model.} We include the results obtained with models trained for cross-validation, using PaiNN (squares), REANN (circles), MACE (triangles), and ACE (stars). Data points and error bars correspond to the RMSE average and standard deviation over 5-fold cross-validation splits. Red markers indicate the final models evaluated using CPU and black markers indicate the final MACE-JAX model evaluated using GPU. The energy and force CPU evaluation times were calculated using a single AMD EPYC 7742 (Rome) 2.25 GHz processor core. The striped circle corresponds to the REANN model which is evaluated on the CPU using the new Fortran-based neighbor list code (REANN-Fort). The GPU evaluation times were evaluated with the NVIDIA V100 GPU.}
        \label{fig:error_vs_time_gpu_si}
    \end{figure}
    Evaluation using GPUs may be employed with the presented models. Evaluating reaction probabilities for dissociative chemisorption requires running many thousands of relatively short trajectories. In this case, employing highly-parallelized CPU architecture is usually more efficient, despite slower evaluation times. However, in some cases, e.g. for larger systems containing hundreds or thousands of atoms, the use of GPUs may be required. In Fig.~\ref{fig:error_vs_time_gpu_si} we compare the RMSEs obtained with our final models evaluated on CPUs with the MACE-JAX models evaluated on GPUs. The MACE-JAX model evaluated on GPUs provides a time-to-solution that is 45 times lower than the final CPU-evaluated MACE models. The sub-linear scaling that MACE-JAX evaluated on GPUs provides with the number of atoms may be crucial to tackle larger systems.

\newpage    
\section{Minimum energy paths} \label{sec:si_meps}

    \begin{figure*}
        \centering
        \includegraphics[width=6.2in]{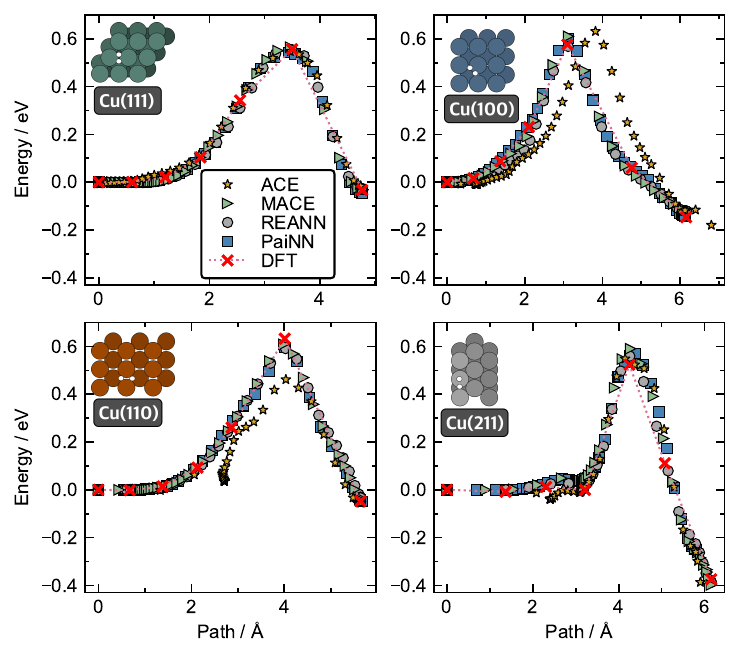}
        \caption{\textbf{Minimum energy paths obtained using CI-NEB method for H\textsubscript{2} dissociative adsorption on different copper surfaces.} Potential energy values are shown along the reaction path ($\textrm{\AA}$), evaluated using DFT (red \textbf{$\pmb{\times}$}) and the MLIPs included in our study: ACE (yellow stars), MACE (green triangles), REANN (gray circles) and PaiNN (blue squares). Figures that show a top-down view of the systems at the transition states are included on the respective plots.}
        \label{fig:meps}
    \end{figure*}
    
    Minimum energy paths (MEP) were obtained using climbing image nudged elastic bands (CI-NEB) for all four Cu facets employing PaiNN, ACE, MACE, and REANN models and DFT, as shown in Fig.~\ref{fig:meps}. The MEPs obtained with DFT match the MEPs obtained with PaiNN, MACE, and REANN models very well, suggesting that barriers are modeled effectively by these models. The MEPs generated with ACE models slightly differ from the models obtained using DFT, with some visible local artifacts, however, the energy barriers are generally similar to the reference values. A substantial difference is observed in the final structures.
    7 images were used for DFT and 50 images were used for all the ML models in CI-NEB calculations with the maximum force along the path of 0.01~eV/$\textrm{\AA}$.

\newpage    
\section{Phonon bandstructure} \label{sec:si_phonon}

    \begin{figure*}
        \centering
        \includegraphics[width=0.48\linewidth]{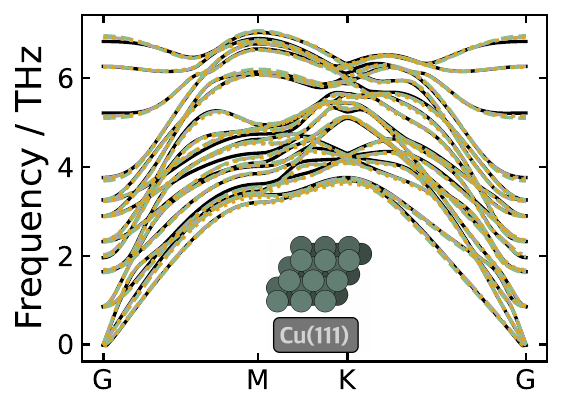}
        \includegraphics[width=0.48\linewidth]{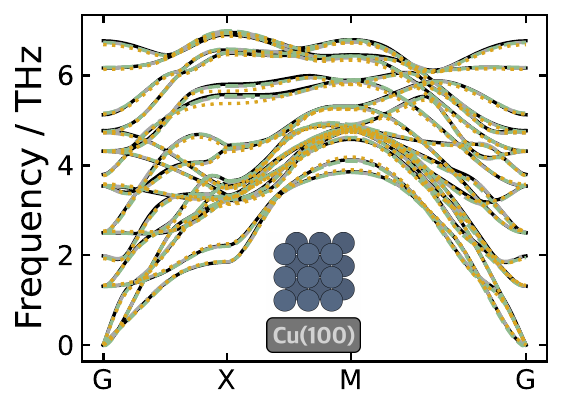}
        \includegraphics[width=0.48\linewidth]{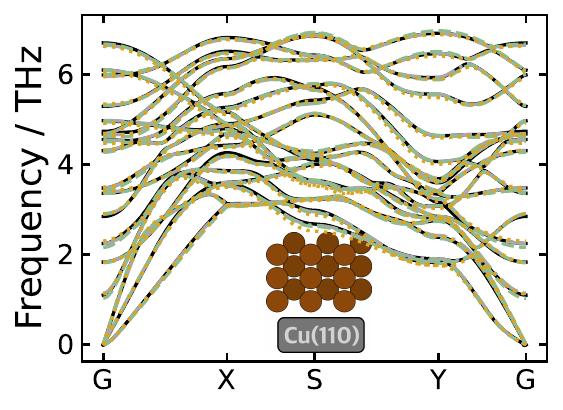}
        \includegraphics[width=0.48\linewidth]{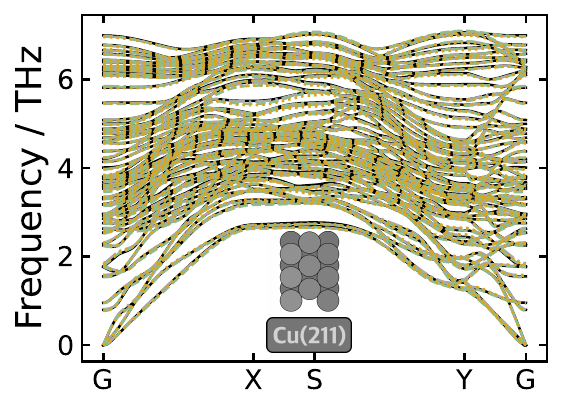}
        \caption{\textbf{Phonon bandstructures for different Cu facets.}  Phonon bandstructures of Cu(111), Cu(100), Cu(110), and Cu(211) surfaces, calculated using ACE~(yellow, dotted lines), MACE~(green, dash-dotted lines), and REANN~(gray, dashed lines) models, compared to the DFT reference calculations (black, solid lines).}
        \label{fig:phonons}
    \end{figure*}

The phonon band structures for Cu(111), (100), (110), and (211) are shown in Fig.~\ref{fig:phonons}. They are evaluated with different models, including ACE, MACE, and REANN, and compared to the DFT-based results. The agreement between MLIPs and DFT is remarkable for all the surfaces, with only minor differences. The bandstructures generated with PaiNN models are not included due to the incompatibility of the PaiNN model with our phonopy/FHI-vibes interface and the presence of negative frequencies throughout the spectra.